\documentclass[a4paper, amsfonts, amssymb, amsmath, reprint, showkeys, nofootinbib, twoside, superscriptaddress]{revtex4-1}
\usepackage[english]{babel}
\usepackage[utf8]{inputenc}
\usepackage[colorinlistoftodos, color=green!40, prependcaption]{todonotes}
\usepackage{amsthm}
\usepackage{mathtools}
\usepackage{physics}
\usepackage{xcolor}
\usepackage{graphicx}
\usepackage[left=23mm,right=13mm,top=35mm,columnsep=15pt]{geometry} 
\usepackage{adjustbox}
\usepackage{placeins}
\usepackage[T1]{fontenc}
\usepackage{lipsum}
\usepackage{csquotes}
\usepackage{xcolor}
\usepackage[pdftex, pdftitle={Article}, pdfauthor={Author}]{hyperref} 
\begin{document}
\title{Shake-up and shake-off spectra in the electron capture decay of atomic $^7$Be}

\author{Mauro Guerra}
  \email[Correspondence email address: ]{mguerra@fct.unl.pt}
  \affiliation{LIBPhys, LA-REAL, NOVA FCT, Universidade NOVA de Lisboa, 2829-516 Caparica, Portugal}

\author{Inwook Kim}
  \affiliation{Lawrence Livermore National Laboratory, Livermore, CA 94550, USA}

\author{Stephan Friedrich}
  \affiliation{Lawrence Livermore National Laboratory, Livermore, CA 94550, USA}

\author{Pedro Amaro}
  \affiliation{LIBPhys, LA-REAL, NOVA FCT, Universidade NOVA de Lisboa, 2829-516 Caparica, Portugal}

\author{Adrien Andoche}
  \affiliation{Université Paris-Saclay, CEA, List, Laboratoire National Henri Becquerel (LNE-LNHB), 91120 Palaiseau, France}

\author{Gonçalo Baptista}
  \affiliation{Laboratoire Kastler Brossel, Sorbonne Université, Paris, France}

\author{Connor Bray}
  \affiliation{Department of Physics, Colorado School of Mines, Golden, CO 80401, USA}

\author{Robin Cantor}
  \affiliation{STAR Cryoelectronics LLC, Santa Fe, NM 87508, USA}

\author{David Diercks}
  \affiliation{Shared Instrumentation Facility, Colorado School of Mines, Golden, CO 80401, USA}
  
\author{Spencer L. Fretwell}
  \affiliation{Department of Physics, Colorado School of Mines, Golden, CO 80401, USA}

\author{Abigail Gillespie}
  \affiliation{Department of Physics, Colorado School of Mines, Golden, CO 80401, USA}

\author{Ad Hall}
  \affiliation{STAR Cryoelectronics LLC, Santa Fe, NM 87508, USA}

\author{Cameron N. Harris}
  \affiliation{Ciambrone Radiochemistry Lab, Patrick Space Force Base, FL 32925, USA}

\author{Jackson T. Harris}
  \affiliation{XIA LLC, Oakland, CA 94601, USA}

\author{Leendert M. Hayen}
  \affiliation{LPC Caen, ENSICAEN, Université de Caen, CNRS/IN2P3, 14000 Caen, France}

\author{Paul Antoine Hervieux}
  \affiliation{Université de Strasbourg, CNRS, Institut de Physique et Chimie des Matériaux de Strasbourg, UMR 7504, F-67000 Strasbourg, France}

\author{Paul Indelicato}
  \affiliation{Laboratoire Kastler Brossel, Sorbonne Université, Paris, France}

\author{Geon Bo Kim}
  \affiliation{Lawrence Livermore National Laboratory, Livermore, CA 94550, USA}

\author{Kyle G. Leach}
  \affiliation{Department of Physics and Astronomy, McMaster University, Hamilton, Ontario L8S 4M1, Canada}

\author{Annika Lennarz}
  \affiliation{TRIUMF, Vancouver, BC V6T 2A3, Canada}

\author{Vincenzo Lordi}
  \affiliation{Lawrence Livermore National Laboratory, Livermore, CA 94550, USA}

\author{Peter Machule}
  \affiliation{Pacific Northwest National Laboratory, Richland, WA 99354, USA}

\author{Andrew Marino}
  \affiliation{Department of Physics, Colorado School of Mines, Golden, CO 80401, USA}

\author{David McKeen}
  \affiliation{Facility for Rare Isotope Beams, Michigan State University, East Lansing, MI 48824, USA}

\author{Xavier Mougeot}
  \affiliation{Université Paris-Saclay, CEA, List, Laboratoire National Henri Becquerel (LNE-LNHB), 91120 Palaiseau, France}

\author{Daniel Pinheiro}
  \affiliation{LIBPhys, LA-REAL, NOVA FCT, Universidade NOVA de Lisboa, 2829-516 Caparica, Portugal}

\author{Francisco Ponce}
  \affiliation{Pacific Northwest National Laboratory, Richland, WA 99354, USA}

\author{Chris Ruiz}
  \affiliation{TRIUMF, Vancouver, BC V6T 2A3, Canada}

\author{Amit Samanta}
  \affiliation{Lawrence Livermore National Laboratory, Livermore, CA 94550, USA}

\author{José Paulo Santos}
  \affiliation{LIBPhys, LA-REAL, NOVA FCT, Universidade NOVA de Lisboa, 2829-516 Caparica, Portugal}

\author{Joseph Smolsky}
  \affiliation{Department of Physics, Colorado School of Mines, Golden, CO 80401, USA}

\author{Caitlyn Stone-Whitehead}
  \affiliation{Department of Physics, Colorado School of Mines, Golden, CO 80401, USA}

\author{Joseph Templet}
  \affiliation{Department of Physics, Colorado School of Mines, Golden, CO 80401, USA}

\author{William K. Warburton}
  \affiliation{XIA LLC, Oakland, CA 94601, USA}

\author{Benjamin Waters}
  \affiliation{Maybell Quantum, Denver, CO 80221, USA}

\author{Jorge Machado}
\email[Correspondence email address: ]{jfd.machado@fct.unl.pt}
  \affiliation{LIBPhys, LA-REAL, NOVA FCT, Universidade NOVA de Lisboa, 2829-516 Caparica, Portugal}

\date{\today} 

\begin{abstract}

The most stringent laboratory-based experimental limits on the existence of sub-MeV sterile neutrinos are currently set by decay spectroscopy of radioactive $^7$Be embedded into superconducting sensors. The systematic uncertainties are dominated by the modeling of the electron shake-up and shake-off spectra that are not based on state-of-the-art atomic theory and do not include electron correlations or relativistic effects. We have used the multiconfiguration Dirac-Fock formalism to obtain correlated wavefunctions \textit{ab initio} and compute all single and double shake processes in the electron capture decay of atomic $^7$Be. The simulations can explain some but not all of the observed spectral features, likely because the wave functions are modified by the Ta sensor material that the $^7$Be is embedded into. The new models also show that the L/K electron capture ratio of $^7$Be in Ta has previously been slightly underestimated revising the previous value of $0.070(7)$ to a new value of $0.0756(20)$.
\end{abstract}


\maketitle

\section{Introduction} \label{sec:outline}

The standard model (SM) of particle physics is an immensely successful framework for understanding the fundamental constituents of the Universe and their interactions. Yet, in spite of its many achievements, it does not include gravity and cannot account for dark matter or the baryon asymmetry in the Universe \cite{1427}. The neutrino sector may be the most promising area to search for physics beyond the standard model (BSM) since neutrino masses are not accounted for in minimal versions of the SM \cite{1441}.
Some BSM physics models postulate the existence of right-handed sterile neutrinos ($\nu_S$) that are inactive in the weak interactions and only interact through gravity and mixing and could be of nearly any mass scale \cite{1428}. 


The Beryllium Electron capture in Superconducting Tunnel junctions (BeEST) experiment currently sets the most stringent limits on the existence of sterile neutrinos in the sub-MeV mass range \cite{1345}. In this experiment, radioactive $^7$Be, with a half-life of $53.284 \pm 0.016$ days \cite{1452}, is directly implanted into high-resolution superconducting tunnel junction (STJ) sensors. When $^7$Be decays by electron capture (EC), the neutrino escapes but the energy of the recoiling $^7$Li daughter can be measured accurately to infer the neutrino mass. The pure EC decaying nucleus of $^7$Be is well-suited for neutrino studies via recoil measurements due to its large Q$_\text{EC}$-value (862 keV) \cite{1453}, relatively high recoil energy ($\sim$57 eV), and simple atomic and nuclear structure. 
In this experiment, radioactive $^7$Be is implanted into high-rate STJs at the TRIUMF-ISAC facility. The sensors are then shipped to Lawrence Livermore National Laboratory (LLNL) to measure the decay products from $^7$Be EC with high energy resolution (1-2 eV). The signal consists of the $^7$Li recoil energy plus the relaxation of the atomic shell from the captured electron or from other excited states created by shake-up and shake-off channels.
Since $^7$Be can capture an electron from its K or L shell, and the $^7$Li daughter nucleus can be produced in its ground or first excited state, the total spectrum due to emission of active neutrinos has four primary peaks: One for K capture and decay into the nuclear ground state of $^7$Li (K-GS), one for K capture and decay into the nuclear excited state $^7$Li$^*$ (K-ES) and the two corresponding L capture peaks L-GS and L-ES. Heavy sterile neutrinos would reduce the $^7$Li recoil energy and provide additional peaks in the spectrum as a signature. 

The sensitivity of the BeEST experiment to $\nu_S$ is determined by the accuracy of the fit to the total spectrum from decays with active neutrinos. The systematic uncertainties of these fits are currently dominated by the functions used to describe electron shake-up (SU) and shake-off (SO) into bound and unbound states upon $^7$Be decay. These effects produce additional peaks in the spectrum and high-energy tails above all peaks since there is an excess energy arising from the excitation (SU) or ionization (SO) of the atomic lithium after electron capture, that will be deposited within the detector. Earlier fits in the BeEST experiment have modeled the SO tails using functions that do not include relativistic and many-electron effects and that are therefore not suitable for a high-precision experiment. In addition, fits to the measured K-GS peak require three separate components, and both the K-GS peaks and the L-GS peak are broadened well beyond the energy resolution of the STJ sensor \cite{1459}. An accurate description of SU and SO is also important for other BSM physics searches. For example, SU and SO loosen the need for resonances in neutrinoless double electron capture and lead to a significant increase of the capture rate in non-resonance nuclei \cite{1445}. Similarly, they affect measurements of the neutrino mass. For example, in the electron capture decay of $^{163}$Ho, a 10\% uncertainty in the Ho-Dy wavefunctions overlap can change the SO probabilities by two orders of magnitude \cite{1447}.

In this work, we have computed the atomic structure of $^7$Be and all relevant levels of the $^7$Li daughter atom and its singly and doubly ionized ions within the multiconfiguration Dirac-Fock (MCDF) formalism. We then calculate the SU and SO probabilities from the overlap between the wave functions of the parent and the daughter atom. The probability that the electron remains in its initial state without change in quantum numbers is used to extract the total shake probability and normalize the spectra. All calculations were performed in the sudden approximation \cite{1430}, assuming only monopole transitions. Additionally, we have included electron correlations in our multiconfiguration Dirac-Fock calculations by augmenting the basis space up to the $4s$ orbital using single and double excitations. 
Finally, since the BeEST experiment is providing an increasingly high-statistics dataset, we also computed the rates of low-probability events such as double SU and double SO.
The following sections describe our \textit{ab initio} calculations and the analytical approximations to the results in the context of the BeEST experiment. We use the results to provide better background modeling in the search for sterile neutrinos in the sub-MeV mass range and to extract a more precise value of the L/K capture ratio of $^7$Be in Ta.
\section{Theoretical methods} \label{sec:develop}

The calculations performed in this work employed the MCDF framework, operating in the \textit{ab initio} Multiconfiguration Dirac-Fock General Matrix Elements (MCDFGME) code developed by Desclaux and Indelicato \cite{92,93}. In MCDFGME, the starting point is the so called \textit{no-pair} Hamiltonian, $\mathcal{H}^{np}$, where projection operators are used to effectively project the electron-electron interaction onto the $E>mc^2$ continuum,

\begin{equation}
    \mathcal{H}^{np}=\sum_{i=1}^{N_e} \mathcal{H}_D(r_i)+\sum_{i<j}\Lambda^{++}_{ij}V_{ij}\Lambda^{++}_{ij}.
\end{equation}
Here, $\mathcal{H}_D(r_i)$ is the one-electron Dirac Hamiltonian, 
and $V_{ij}$ represents the electron-electron interaction through some suitable operator, (e.g., Coulomb or Breit interaction). 

The projection operator $\Lambda^{++}_{ij}=P^+_iP^+_j$ is used in one-electron eigenfunctions to project them into positive energy wavefunctions, avoiding the coupling of positive and negative energy continua. The Coulomb-Breit operator can be written as

\begin{eqnarray}
V_{ij}&=&\frac{1}{r_{ij}}-\frac{\boldsymbol{\alpha}_i\cdot\boldsymbol{\alpha}_j}{r_{ij}}\cos(\omega_{ij}r_{ij})\nonumber\\
&+&(\boldsymbol{\alpha}_i\cdot\boldsymbol{\nabla}_i)(\boldsymbol{\alpha}_j\cdot\boldsymbol{\nabla}_j)\frac{\cos(\omega_{ij}r_{ij})-1}{\omega_{ij}^2r_{ij}},\label{potential}
\end{eqnarray}
where $r_{ij}=(\boldsymbol{r}_i-\boldsymbol{r}_j)$ is the interelectronic distance, $\omega_{ij}$ is the energy of the photon exchanged between the two electrons, and $\boldsymbol{\alpha}_i$ are Dirac matrices. 

The first term, $1/r_{ij}$, describes the usual Coulomb interaction, the second is the Gaunt interaction, and the remaining term is the lowest-order retardation interaction due to the finite speed of light. These potentials are included in the self-consistent field calculation, while the remaining Breit retardation terms are included as perturbations. Quantum ElectroDynamics (QED) radiative corrections, such as self-energy and vacuum polarization, are also included in the calculations \cite{1388}. The one-electron self-energy is computed using the expressions proposed by Mohr and Kim \cite{1409,1410} and further corrected for finite nuclear size \cite{1411}. The Welton formalism is used to compute the self-energy screening due to the remaining electrons \cite{1412}. Regarding the vacuum polarization, the Uelhing potential is included self-consistently to all orders, and the Wichmann-Kroll and Kallen-Sabry contributions are computed as perturbations \cite{512}. The latest version of the MCDFGME code (V2025v1) already implements to all-order the Wichmann and Kroll potential \cite{1465}.
The QED computation accuracy for few electron atoms and ions can be seen in \cite{1466}.
Nuclear size effects are computed considering the Thomas Fermi nuclear model, and the measured values for atomic masses and charge radii are taken from the tables of Audi \textit{et al.} \cite{1413} and Angeli \cite{1414}, respectively.

\subsection{Relativistic State functions}

Wavefunctions in the multiconfiguration Dirac-Fock method are calculated from the variational principle and constructed from linear combinations of Configuration State Functions (CSF). In turn, these CSF are themselves expressed as linear combinations of Slater determinants of Dirac spinors,

\begin{eqnarray}
&&\Phi_{\Pi J M}^{\eta}(\vec{r}_1,\cdots,\vec{r}_{N_e})=\nonumber\\
&&\sum_{q\geq 1} C_q^\eta\left| 
\begin{array}{ccc}
\phi^q_{n_1\kappa_1m_1}(\vec{r}_1) & \cdots & \phi^q_{n_{N_e}\kappa_{N_e}m_{N_e}}(\vec{r}_1) \\ 
\vdots & \cdots & \vdots \\
\phi^q_{n_1\kappa_1m_1}(\vec{r}_{N_e}) & \cdots & \phi^q_{n_{N_e}\kappa_{N_e}m_{N_e}}(\vec{r}_{N_e})
\end{array} \right|,\label{slater}
\end{eqnarray}

\noindent where the four-component Dirac spinors are \cite{512}

\begin{equation}
\phi_{n\kappa m}(r)=\frac{1}{r}\left [ 
\begin{array}{c}
P_{n\kappa}(r)\chi_{\kappa \mu}(\theta, \phi) \\
i Q_{n\kappa}(r)\chi_{-\kappa \mu}(\theta, \phi) \end{array} \right ].
\end{equation}
Here, $\chi_{\kappa \mu}(\theta, \phi)$ is the two-component Pauli spherical spinor, $n$ is the principal quantum number, $\kappa$ is the Dirac quantum number, $m$ is the total angular momentum projection, and $\mu$ is the eigenvalue of $\hat{J_z}$.
The linear combination's coefficients and the number of CSFs are obtained by constraining the solution to be an eigenfunction of the Hamiltonian, $\hat{H}$, the total angular momentum operator $\hat{J^2}$, and projection operator $\hat{J_z}$. All other quantum numbers that are required to characterize the wavefunction are represented by $\eta$ in Eq. \ref{slater}.

The multiconfiguration calculation is a generalization obtained by constructing an Atomic State Function (ASF) as a linear combination of CSFs corresponding to fundamental and excited electronic configurations,
\begin{equation}
\Psi_{\Pi J M}(\vec{r}_1,\cdots,\vec{r}_{N_e})=\sum_{\eta=1}^{N_\text{CSF}}D_\eta\Phi_{\Pi J M}^{\eta}(\vec{r}_1,\cdots,\vec{r}_{N_e}).
\end{equation}

This method is implemented in the MCDFGME code by summing CSFs that account for single and double excitations of the valence electrons up to a certain principal quantum number, $n\geq n_\text{valence}$. This growth in the basis set has a computational cost that scales rapidly with the number of included CSFs, which, in turn, grows rapidly with $n$. This effect is very relevant for atomic calculations, especially for valence electrons in open-shell atomic systems. In fact, even without the inclusion of single and double excitations, each ASF comprises a linear combination of \textit{jj} CSFs for the same angular momentum, since the CSFs are computed in the \textit{jj} coupling scheme and the MCDFGME input is given in terms of \textit{LS} configurations. 
For simplicity, we will describe the wavefunctions, $\Psi_{\Pi J M}(\vec{r}_1,\cdots,\vec{r}_{N_e})$, with their $n \ell j$ quantum numbers in the following sections, labling them as $\psi(n\ell j)$. The MCDFGME version V2025v1 \cite{1465} used in this work was developed specifically for BeEST since it now allows the computation of shake probabilities between atoms and ions with different nuclear charges.

\subsection{Atomic Shake Processes}

Shake processes in atoms occur when there is a sudden change in the potential felt by the atomic electrons so that their wavefunctions are no longer eigenstates of the new Hamiltonian. 
After the nuclear capture of a K or an L electron, the atom is left in a perturbed state where each of the remaining electrons has a different potential energy than before the capture process. In a few attoseconds there is a probability that one or more electrons are ejected (shaken) from the atomic system while the others relax to the resulting ion's ground state. 
The initial system wavefunctions can be written as an expansion over the new system eigenstates
\begin{equation}
    |\psi_i(n\ell j) \rangle = \sum_{n'\ell' j'} c_{if} |\psi_f(n'\ell' j') \rangle 
\end{equation}
where $c_{if}=\langle \psi_f(n'\ell' j')|\psi_i(n\ell j)\rangle$ are the overlaps between the eigenstates of the initial system and the eigenstates of the system after electron capture. The quantity $|c_{if}|^2$ represents the probability that the system in an initial state $\psi_i(n\ell j)$ ends up in a final state $\psi_f(n'\ell' j')$ after a sudden change in the potential due to nuclear electron capture. This final state can span over all allowed states of the new Hamiltonian as well as continuum wavefunctions.
The process where one or more of the electrons transition to continuum wavefunctions is called shake-off. Alternatively, shake-up describes processes where one or more of the electrons are excited into higher bound states while the other electrons rearrange to a more favorable configuration. The joint probability for both effects is usually calculated within the Sudden Approximation (SA) formalism \cite{1311}.

\begin{figure}[tb]
    \includegraphics[width=\linewidth]{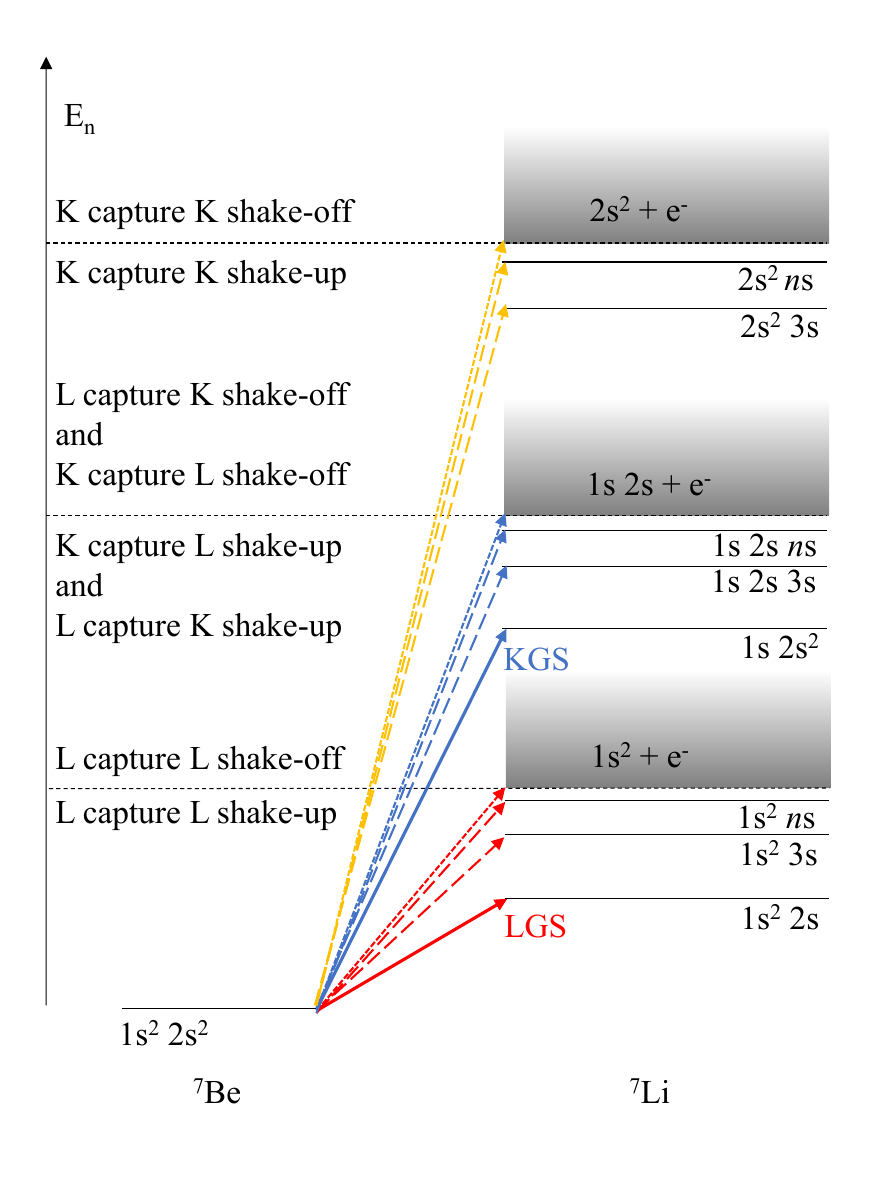}
    \caption{Schematic of the atomic processes after electron capture. The solid lines represent the capture events that result in the two main peaks of the BeEST spectra (K-GS, L-GS) and their excited state counterparts (K-ES, L-ES). The dashed lines represent shake-up channels and the dotted lines represent the onset of K and L electron shake-off after either K or L capture.}
    \label{fig:scheme}
\end{figure}

In the BeEST experiment, both SU and SO can occur with different probabilities after electron capture in $^7$Be, and the energy necessary for each of the processes reduces the energy of the emitted neutrino. Given the high $Q_{\mathrm{EC}}$ value of the $^7$Be decay and momentum conservation, the change in the nuclear recoil energy is negligible. The SU and SO processes in $^7$Be EC are summarized schematically in Fig. \ref{fig:scheme}. All of the processes are represented by arrows, starting at the $^7$Be ground state and ending in a $^7$Li state after recombination inside the detector. In the right column of Fig. \ref{fig:scheme} we present as solid lines the intermediate states arising from simple capture ($1s^2 2s$ and $1s 2s^2$ for L and K capture, respectively) and SU channels where one of the K or L electrons is shaken up after either K or L capture. SO processes, where a K or L electron is ejected from the atom after K or L capture, are represented in gray. All of the processes are labeled in the left column at the energy region where they occur.

\subsection{Total Shake Probability}

The total shake probability, $P_{n\ell j}^{\text{shake}}$, for an electron characterized by quantum numbers $\qty(n\ell j)$ in the sudden approximation (SA) is calculated by subtracting from the total probability space the likelihood that all the $N_{n\ell j}$ electrons in a shell retain their quantum numbers after the decay. 

\begin{equation}
    P_{n\ell j}^{\text{shake}}=1-\qty| \braket{\psi_{f\,\qty(n\ell j)}}{\psi_{i\,\qty(n\ell j)}}|^{2 N_{n \ell j}}-N_{n \ell j} P_{n \ell j}^{\text{forb}}, \label{shake}
\end{equation}

\noindent where


\begin{equation}
\label{eq:forb}
    P_{n \ell j}^{\text{forb}}= \sum_{\qty(n'\ell j)\neq \qty(n\ell j)} \frac{N_{n' \ell j}}{2j +1} \qty(\qty| \braket{\psi_{f\,\qty(n'\ell j)}}{\psi_{i\,\qty(n\ell j)}}|^{2} ).
\end{equation}

\noindent Here $\psi_{f\,\qty(n'\ell j)}$ are shells that are already fully or partially occupied and thus $P_{\text{forb}}$ considers the probability of transitions to levels forbidden by the Pauli exclusion principle that has to be removed from the total space of probabilities. The quantities $N_{n'\ell j}$ and $N_{n\ell j}$ refer to the corresponding post-capture shell occupancies \cite{1363}. This differs from the Carlson--Nestor convention, where neutral-atom occupancies are used \cite{1311}. In this way, the shake probabilities represent a fraction of all atomic channels arising from the same EC event. Hence, a comparison to the experiment requires disentangling all atomic processes in the fitting procedure, and the sum of all processes from the same capture event should be normalized to one.

The probability in Eq. \ref{shake} has been interpreted many times, erroneously, as the probability of the excitation of \textit{one} electron, either into another bound state or into the continuum, after a sudden change in the atomic potential \cite{1430}. In fact, it represents the probability that \textit{one or more} of the electrons are shaken up or off from a given orbital. Although the shake probability for one electron is usually much higher than multiple electrons being shaken \cite{1419}, double and triple shake can amount in some atomic systems to 10-30\% of the total shake probability. Thus, these effects should not be thought of as higher-order corrections in weak nuclear decay, but leading order terms in the sudden approximation.
Multiple shake probabilities can be calculated from the total shake values, as it is a similar problem as calculating probabilities of any number of successes from a number of independent trials. For the probability of a single electron being removed from any shell, we can use binomial probability by making the assumption that each of the $N_{n\ell}$ electrons in a shell is equally likely to be ejected as a result of the change in the Hamiltonian. Thus, multiple shake probabilities can be calculated from \cite{1363}

\begin{multline}
\label{eq:multishake}
    P_{n\ell j}(n_e)= \binom{N_{n\ell j}}{n_e} \qty(1-\qty(1-P_{\text{shake}})^{\frac{1}{N_{n \ell j}}})^{n_e} \\
    \times \qty(1-P_{\text{shake}})^{\frac{N_{n \ell j}-n_e}{N_{n \ell j}}},
\end{multline}

\noindent where $n_e$ is the number of shaken electrons from shell $N_{n \ell j}$ and $P_{\text{shake}}$ is the probability that \textit{one or more} of the electrons are shaken from that shell (Eq. \ref{shake}). This expression can be used to compute single shake probabilities or multiple shake from the same orbital. Double shake probabilities with electrons from different shells are calculated following the procedure in \cite{1363}, where the double shake probability is computed as the product of single shake probabilities from the shells involved in the transition.

\subsection{Shake-up}

The probability that electron capture simultaneously promotes another electron to an excited level can be computed in the SA from first principles. This probability is merely the squared modulus of the overlap between the initial wavefunction before electron capture and the final wavefunction after electron capture. Since this process in the SA regime is treated as a two-step process, these excitations must retain the parity of the atomic wavefunction. Hence, if an electron can be described by the wavefunction $\psi_{i\,\qty(n\ell j)}$, the probability that any of the $N_{n\ell j}$ electrons with this wavefunction might be excited to another level with the wavefunction $\psi_{f\,\qty(n'\ell j)}$ is given by

\begin{equation}
    P_{n \ell j}^{\text{shake-up}}=N_{n\ell j} \qty| \braket{\psi_{f\,\qty(n'\ell j)}}{\psi_{i\,\qty(n\ell j)}}|^{2}. \label{shakeup}
\end{equation}

\subsection{Shake-off} \label{sec:shakeoff}

SO probabilities are calculated similar to SU \ref{shakeup}, except that in this case continuum electron wavefunctions are used as final wavefunctions. The process involves calculating the integral of the differential probability across an infinite number of energy levels of free electrons, up to the reaction $Q_{EC}$ value. Given that the normalization of the continuum wavefunctions is energy dependent, only the shape of the SO distribution can be computed as a function of the electron energy.
However, if the total shake and SU probabilities have already been determined, the SO total probability can be obtained by subtracting the contributions for single and multiple SU transitions from the total shake probability. This value represents the total probability for SO of one or more electrons, effectively allowing the calculation of the relative area of the SO distribution from first principles. 

\begin{figure}[tb]
    \centering
    \includegraphics[width=\linewidth]{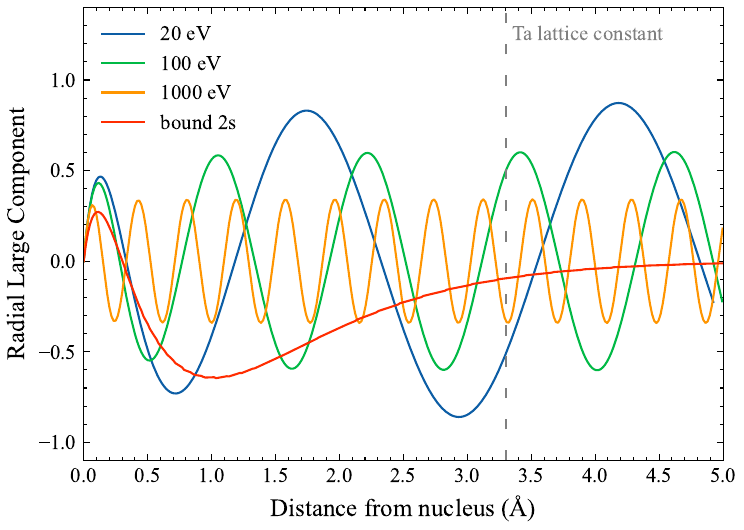}
    \caption{Radial wavefunction of a bound $2s$ electron in the $^7$Be ground state $1s^2 2s^2$ and continuum wavefunctions (computed in the $^7$Li$^+$ potential) of an electron with kinetic energies of 20 eV, 100 eV or 1000 eV after a ground state L capture, $1s^2 + e^-$. The gray dashed line at $3.3$ \AA~is the lattice constant of the Ta host in the BeEST experiment.}
    \label{fig:shakeoff}
\end{figure}

The functional form of these probabilities can be computed as a function of the energy in a similar fashion to the SU probabilities. For $^7$Be EC, the SO spectra are computed from the squared modulus of the overlap between the bound $^7$Be wavefunctions and the continuum wavefunctions for different  kinetic energies of the ejected electron, calculated in the potential of the $^7$Li ion after K or L capture for ejected K or L electrons. To illustrate the method, Fig. \ref{fig:shakeoff} shows the radial wavefunctions of the $2s$ bound electron in $^7$Be and the continuum electron wavefunctions for three kinetic energies in the potential of a $^7$Li ion with a capture hole in the K shell and a $2s$ hole from the ejected L electron. As expected, given the oscillatory nature of the continuum wavefunctions, the overlap decreases with increasing kinetic energies, resulting in a specific profile of the SO spectra. Since this shape is driven essentially by the wavefunction of the bound electron, the profiles will be different for K and L SO.
Thus, in this particular experiment, we will have four different profiles arising from the four SO channels: $1s^2 2s^2 \rightarrow 2s^2 + e^-$ which corresponds to a K capture K SO, $1s^2 2s^2 \rightarrow 1s 2s + e^-$ as a K capture L SO, $1s^2 2s^2 \rightarrow 1s 2s + e^-$ as L capture K SO and $1s^2 2s^2 \rightarrow 1s^2 + e^-$ corresponding to an L capture L SO. The gray dashed line in Fig. \ref{fig:shakeoff} represents the Ta lattice constant of $3.3$ \AA~to show the range to which the continuum wavefunctions are distorted by the daughter ion potential and compare it to the relative position of the other atoms in the Ta matrix. Other quantities of interest are the Ta covalent radius of 1.46 \AA~ \cite{1030} and its atomic radius, which range from 1.55 \AA~\cite{1153} to 2.00 \AA~\cite{38} depending on whether relativistic wavefunctions are used or not to compute this value.
All SO transitions are illustrated in Fig.\ref{fig:scheme}. 

\section{Results} \label{sec:results}
Accurate theoretical modeling of level energies and SU/SO probabilities is essential for interpreting the BeEST spectra. These effects, originating from the sudden change of the nuclear potential during electron capture, can lead to electronic excitations (SU) or ionizations (SO), thereby redistributing spectral strength into satellite channels. A reliable treatment of such contributions is therefore indispensable both to benchmark the experimental data and to establish robust limits on new physics.

\subsection{Level Energies}
\begin{figure}[tb]
    \includegraphics[width=\linewidth]{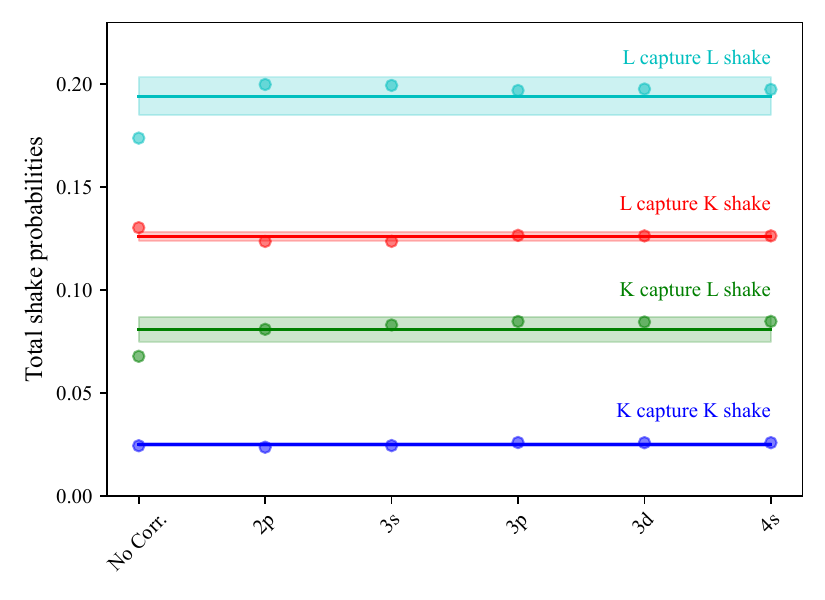}
    \caption{Total shake probabilities according to Eq. (\ref{shake}) as a function of the size of the basis space. All the single and double excitations up to the $4s$ orbital are included. The colored bands represent the $1\sigma$ intervals.}
    \label{fig:corr}
\end{figure}

The accuracy of the energies depends on the size of the basis space in the multiconfiguration Dirac-Fock calculations. Here, we have included electron correlations by augmenting the basis space up to the 4s orbital using single and double excitations. Although higher accuracy is possible \cite{1415,1417,1418}, there are limited benefits to improvements beyond the ~0.1 eV accuracy of the BeEST experiment \cite{1459}, especially since we expect the Ta host material to affect the experimental energy values of the $^7$Be spectrum \cite{1408}. All calculations were performed for mass-7 nuclei using a Fermi distribution. Nuclear radii for $^7$Be and $^7$Li were taken from Angeli and Marinova ~\cite{1439}.
%
While the inclusion of higher levels of electronic correlation systematically reduces the calculated energies, the values have approached an asymptotic limit once correlations up to the 4s orbital are included. We found a similar evolution with the number of correlated orbitals for the SU and SO processes (Fig.~\ref{fig:corr}). Therefore, we report values calculated for a basis space up to the 4s orbital. To estimate the error caused by this approximation, we used an exponential model to extrapolate to the limit of an infinite basis set and used this limit as a measure of the systematic uncertainty in our calculations. We found energies of -147.05(3), -203.38(3) and -52.28(3) eV for the $1s2s^2$, $1s^22s$ and $2s^23s$ configurations in Li, respectively. These values are within 2$\sigma$ to 3$\sigma$ of the most accurate current calculations~\cite{1435,1437}. They predict a separation of 56.33(5) eV between the K-GS and the L-GS peak, and 93.15(5) eV between the K-GS and the K-GS K-SU peak for isolated $^7$Li atoms. The experimental values from the BeEST phase III data are significantly lower at 51.57(16) eV and 89.37(27) eV, respectively. 
While these differences may be attributed to solid-state effects and a possible detector non-linearity in response to nuclear recoils, they also provide a valuable opportunity to refine our understanding of environmental influences on the atomic spectrum and to further improve the experimental methodology.

\subsection{Total Shake Probability}
The probability that an electron capture decay excites (``shakes'') one or more of the remaining electrons out of a given shell can be computed from the initial- and final-state wavefunctions in the sudden approximation according to Eqs. (\ref{shake}) and (\ref{eq:forb}). For the $^7$Be EC decay, it includes all K capture decays that do not produce $^7$Li in a final $1s2s^2$ configuration and all L capture decays to configurations other than $1s^22s$. Calculations of the total shake probability therefore require the 1s and 2s wavefunctions for the ground state of $^7$Be ($1s^22s^2\;^1S_0$) and $^7$Li ($1s^22s\;^2S_{1/2}$) and the wavefunction of the $^7$Li excited state ($1s2s^2\;^2S_{1/2}$). We have calculated the total capture possibilities using Eqs. (\ref{shake}), (\ref{eq:forb})  and (\ref{eq:multishake}) for different sizes of the basis space using CSFs with single and double excitations up to the 4s orbital. In all but the L capture K shake processes, the correlated wavefunctions show an increasing overlap that cause the shake probabilities to increase slightly compared to calculations without electron correlations. This effect has recently also been observed in the multiconfiguration Dirac-Hartree-Fock calculations of Nguyen et al.~\cite{1363} for Cu. It is still unclear why we observe a decrease of the shake probability for the L capture K shake process of around 0.4\%, even if it is only 1.5 times our estimated uncertainty. The shake probabilities are sensitive to the wavefunction shape, and the convergence of the shake probabilities does not show a trend similar to that of the energies. We therefore conservatively estimated the uncertainty for total shake probabilities from the standard deviation of the shake results for different numbers of CSFs included in the correlated wavefunctions (Fig. \ref{fig:corr}).
Double-electron KL shakes were obtained by multiplying the single shake probability for K and L shakes after K or L capture as described in \cite{1363}. Uncertainties are computed by error propagation of the standard deviation of the probability for different levels of correlation. The single-electron, two-electron and total shake probabilities for the electron capture of $^7$Be are summarized in Table~\ref{tab:shake}.

\begin{table*}[tb]
\centering
\caption{Single-shake, double-shake and total shake probabilities for K- and L-capture decays of $^7$Be. All values are computed relative to the capture events so that the total shake probabilities plus the no-shake probability sum to 1. Double KK and LL shake probabilities are computed directly with Eq. \ref{eq:multishake} while KL shake probabilities are computed from the conditional probability of a single K and single L shake probabilities.}
\begin{tabular}{ l c c l c}
\multicolumn{5}{c}{Shake probabilities} \\
Process & Total shake & Single shake & Process & Double shake \\ 
\hline
K capture K shake	&	$0.0260(9) $ & $0.0258(9) $ & K capture KL shake & $0.0021(2) $ \\
K capture L shake	&	$0.085(6) $ & $0.083(6) $ & K capture LL shake & $0.0019(1) $ \\
L capture K shake	&	$0.126(3) $ & $0.122(2) $ & L capture KK shake & $0.00427(8) $ \\
L capture L shake	&	$0.198(9) $ & $0.187(9) $ & L capture KL shake & $0.023(1) $
\end{tabular}
    \label{tab:shake}
\end{table*}

\subsection{Shake-Up}

\begin{figure}[h]
    \centering
    \includegraphics[width=\linewidth]{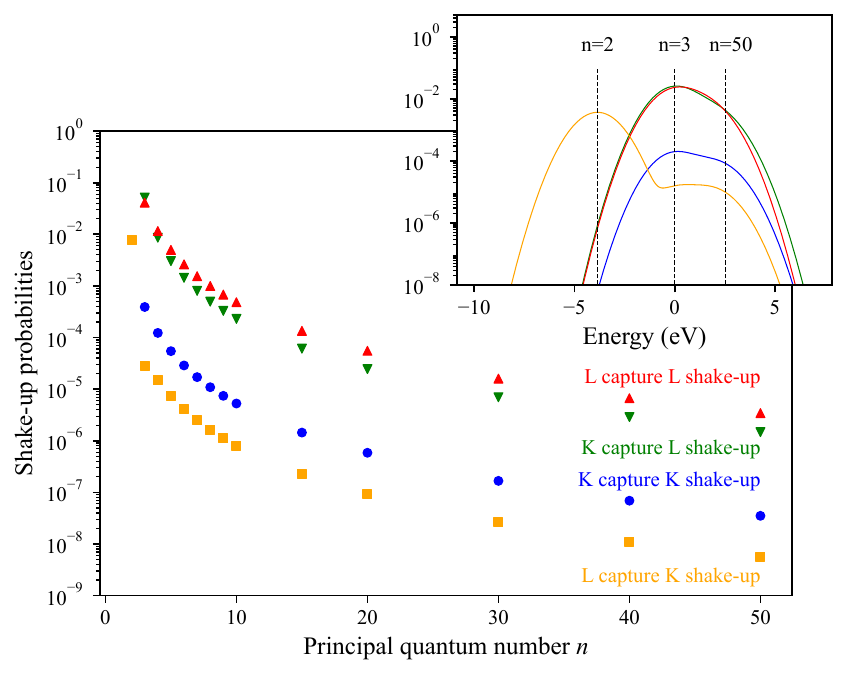}
    \caption{Shake-up probability for $1s\rightarrow ns$ and $2s\rightarrow ns$ electron excitations after K or L shell capture in $^7$Be. The inset shows the sum of all of the SU channels with 2 eV FWHM Gaussian broadening to account for the detector resolution. Energies are shown relative to the $n=3$ SU channel of each dataset.}
    \label{fig:shakeup}
\end{figure}

The electron SU probabilities into empty bound states of $^7$Li were computed using Eq.~(\ref{shakeup}) since it directly provides the probability that the atom ends up in a given eigenstate of the new Hamiltonian after the change in potential due to electron capture \cite{1462}. For isolated Be atoms in the ground state, only transitions into $s$ orbitals are allowed due to conservation of the total angular momentum, and SU is dominated by transitions into the lowest unoccupied 3s orbital (Fig. \ref{fig:shakeup}). For L SU, this probability is several percent of the total (Table~\ref{tab:shakeup}). 
The SU probabilities tend to decrease rapidly as the principal quantum number increases, after both K and L electron captures (Fig. \ref{fig:shakeup}). To obtain the total SU probability, it is necessary to extrapolate the computed values since, in principle, the electron can be excited to any of the infinite set of orbitals of the daughter ion. In fact, the solid state calculations for the Be in the Ta matrix \cite{1408}, show that the $2s$ orbital is already degenerate with the Fermi level, which means that the shape of the SU peaks will reflect the highly asymmetric distribution of the density of states.
The total SU values for the K and L capture in $^7$Be are obtained by extrapolating the curve fits of the data in Fig. \ref{fig:shakeup} to $n=200$, above which the SU probability is negligible (Table \ref{tab:shakeup}).

The SU peak shapes are given by the sum of all SU states, broadened with a 2 eV Gaussian FWHM to account for the STJ detector resolution.  (Fig. \ref{fig:shakeup}, inset). Aside from the L capture K-SU peak, all SU peaks have a similar asymmetrical structure, with a peak due to SU into the lowest unoccupied $3s$ orbital and a high-energy shoulder due to SU into states with $n>3$. This broadens the SU peaks beyond the detector resolution of 2 eV. In contrast, the L capture K-SU spectrum shows two peaks since the $2s$ orbital is only partially occupied after L capture so that the SU transition $1s^2 2s \rightarrow 1s 2s^2$ is possible. The energy of this transition is equal to that of the K-GS peak, and it should be accounted for when calculating the L/K capture ratio (Sec. IV C). The two peaks are separated by 3.85 eV (Table~\ref{tab:shakeup}). The same energy separates the L-SU peaks from the primary "no-shake" peak, which may explain the need for a secondary component at higher energy in the fit to the K-GS peak. The only other SU transition that is visible experimentally is due to K-SU after K capture as it is well separated from the four primary peaks.

\begin{table*}[tb]
\centering
\caption{Shake-up probabilities after K and L capture. The total is the sum of all SU channels up to $n=200$, calculated from a quadratic interpolation of the presented results. The contribution of SU probabilities from levels with $50<n<200$ is less than $0.02\%$. Shown also, are the ionization thresholds corresponding to the particular SO channel's energies ($\infty$).}
\begin{tabular}{ c c c c c c c c}
 & KC-LSU & LC-KSU & Energy (eV) & KC-KSU & Energy (eV) & LC-LSU & Energy (eV) \\ 
n & \multicolumn{2}{c}{$1s^2 2s^2 \rightarrow 1s~2s~ns$}  & & $1s^2 2s^2 \rightarrow 2s^2 ns$ & & $1s^2 2s^2 \rightarrow 1s^2 ns$ & \\
\hline
2	&		&	7.8(6)E-03	&	113.65(3)	&		&		&		&		\\
3	&	5.2(4)E-02	&	2.8(2)E-05	&	117.50(3)	&	3.9(3)E-04	&	206.80(5)	&	4.1(3)E-02 	&	60.16(3)	\\
4	&	8.6(7)E-03	&	1.5(2)E-05	&	118.83(3)	&	1.23(9)E-04	&	208.27(5)	&	1.16(9)E-02	&	61.12(3)	\\
5	&	3.0(3)E-03	&	7.4(6)E-06	&	119.33(3)	&	5.4(4)E-05	&	208.82(5)	&	4.9(4)E-03	&	61.53(3)	\\
6	&	1.4(2)E-03	&	4.1(3)E-06	&	119.57(3)	&	2.9(3)E-05	&	209.08(5)	&	2.6(2)E-03 	&	61.73(3)	\\
7	&	8.0(6)E-04	&	2.5(2)E-06	&	119.71(3)	&	1.7(2)E-05	&	209.22(5)	&	1.6(2)E-03 	&	61.86(3)	\\
8	&	4.9(4)E-04	&	1.6(2)E-06	&	119.79(3)	&	1.09(8)E-05	&	209.31(5)	&	9.9(7)E-04 	&	61.93(3)	\\
9	&	3.3(3)E-04	&	1.12(8)E-06	&	119.85(3)	&	7.4(6)E-06	&	209.37(5)	&	6.8(5)E-04 	&	61.98(3)	\\
10	&	2.3(2)E-04	&	8.0(6)E-07	&	119.89(3)	&	5.3(4)E-06	&	209.41(5)	&	4.8(4)E-04 	&	62.02(3)	\\
15	&	6.1(5)E-05	&	2.2(2)E-07	&	119.97(3)	&	1.4(2)E-06	&	209.51(5)	&	1.35(10)E-04 	&	62.10(3)	\\
20	&	2.5(2)E-05	&	9.2(7)E-08	&	120.00(3)	&	5.8(5)E-07	&	209.54(5)	&	5.5(4)E-05 	&	62.13(3)	\\
30	&	6.9(5)E-06	&	2.6(2)E-08	&	120.02(3)	&	1.7(2)E-07	&	209.56(5)	&	1.6(2)E-05 	&	62.15(3)	\\
40	&	2.9(3)E-06	&	1.10(8)E-08	&	120.03(3)	&	6.9(5)E-08	&	209.56(5)	&	6.7(5)E-06 	&	62.16(3)	\\
50	&	1.5(2)E-06	&	5.6(4)E-09	&	120.03(3)	&	3.5(3)E-08	&	209.57(5)	&	3.4(3)E-06 	&	62.16(3)	\\
$\infty$ &  &  & 120.04(3) &  & 209.58(5) &  & 62.17(3)  \\
\hline
Total	&	0.067(5) &	0.0079(6) &	& 0.00066(5)	& &	0.067(5) & \\
\end{tabular}
    \label{tab:shakeup}
\end{table*}

\subsection{Shake-Off}

SO probabilities into unbound states are also calculated according to Eq.~(\ref{shakeup}), except that in this case the final state is given by a continuum wavefunction (Sec.~\ref{sec:shakeoff}). We have calculated the SO probabilities for kinetic energies up to 1 keV and extrapolated to $Q_{\mathrm{EC}}$=862 keV. SO spectra are normalized by subtracting the single-electron (Table~\ref{tab:shakeup}) and two-electron SU probabilities (Table~\ref{tab:doubleshakeup}) from the total shake probability (Table~\ref{tab:shake}) \cite{1419}. For K capture, SO is found to occur with a probability of a few percent, while L capture is accompanied by SO in around 25\% of the decays (Table~\ref{tab:shakeoff}). The fact that SO is significantly stronger after L capture is due to a larger change in the effective nuclear charge ($Z_{eff}$) seen by the remaining K and L electrons. The change in $Z_{eff}$ after K capture is small, given that the nuclear charge seen by the L electrons (and to a lesser extent also by the K electrons) remains similar due to charge conservation in EC. In fact, the relative change in effective nuclear charge after K capture is of 22\% for K electrons and 11\% for L electrons, while the change in $Z_{eff}$ after L capture is of 27\% for K electrons and 31\% for L electrons. This effect has been calculated before, and the same behavior was observed in a set of elements ranging from $7<Z<54$ \cite{1444} although it was not fully appreciated for the analysis of the BeEST phase-II spectra~\cite{1368}. Threshold energies for the various single SO spectra reflect the binding energies of the electrons relative to the ionization threshold (Table~\ref{tab:shakeoff})).

\begin{table}[h]
\centering
\caption{Single shake-off probabilities and energy thresholds for K and L electrons after K and L electron capture. These results are obtained by subtracting the sums from Table \ref{tab:shakeup} from the single shake probabilities in Table \ref{tab:shake}.}
\begin{tabular}{ l c c}
Process & Shake-off & Energy threshold \\ 
& probability & (eV) \\
\hline
K capture K shake-off	&	$0.0251(10) $ & 209.58(5) \\
K capture L shake-off	&	$0.016(8) $ & 120.04(4) \\
L capture K shake-off	&	$0.114(4) $ & 120.04(4) \\
L capture L shake-off	&	$0.120(11)  $ & 62.17(3)
\end{tabular}
    \label{tab:shakeoff}
\end{table}

\begin{figure}[tb]
    \centering
    \includegraphics[width=\linewidth]{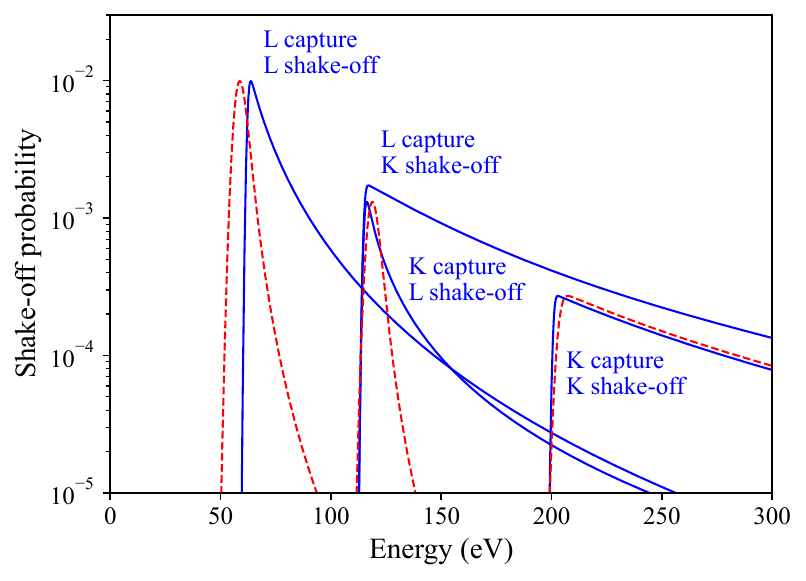}
    \caption{Comparison of the Levinger shake-off profiles used in the BeEST phase-II analysis (red dashed line) and the \textit{ab initio} calculations (blue solid line). All profiles are convolved with a 2 eV FWHM Gaussian to account for the detector resolution. The energy thresholds for the Levinger spectra are the fit values, and the computed values for the \textit{ab initio} calculations.}
    \label{fig:shakeoff_comp}
\end{figure}

It is instructive to compare these spectra with earlier models of the SO process, which have traditionally been based either on exponentially modified Gaussian distributions \cite{1368} or on an approach by Levinger from the 1950s~\cite{1421,1422,1345}. The latter approach used screened hydrogenic wavefunctions and should therefore be sufficiently accurate to describe K SO. These wavefunctions are, however, expected to be less accurate to describe n = 2 states and thus L SO, which depends strongly on the wavefunction shape~\cite{1153}. Figure~\ref{fig:shakeoff_comp} compares our \textit{ab initio} calculations of the SO profiles with those of the BeEST phase-II results based on Levinger’s approach~\cite{1345}. As expected, the K SO spectra are quite similar for the two approaches, although we can only compare them for K capture since the phase-II fit did not include K-SO after L capture. In contrast, the Levinger functions for L SO decay much quicker than the \textit{ab initio} calculations. This is likely because Levinger used nonrelativistic calculations while we have used the relativistic Dirac-Fock formalism, and because Levinger used screened hydrogenic wavefunctions while we use multiconfiguration Dirac-Fock wavefunctions (Sec.~\ref{sec:develop}). Phase-III of the BeEST experiment therefore benefits from greatly improved SO spectra, especially for L SO. The use of exponentially modified Gaussian distributions \cite{1368} and Levinger functions \cite{1345} in BeEST, albeit returning different fit shapes, rendered very similar L/K ratios within their quoted uncertainty.

\begin{figure*}[tb]
    \centering
    \includegraphics[width=\linewidth]{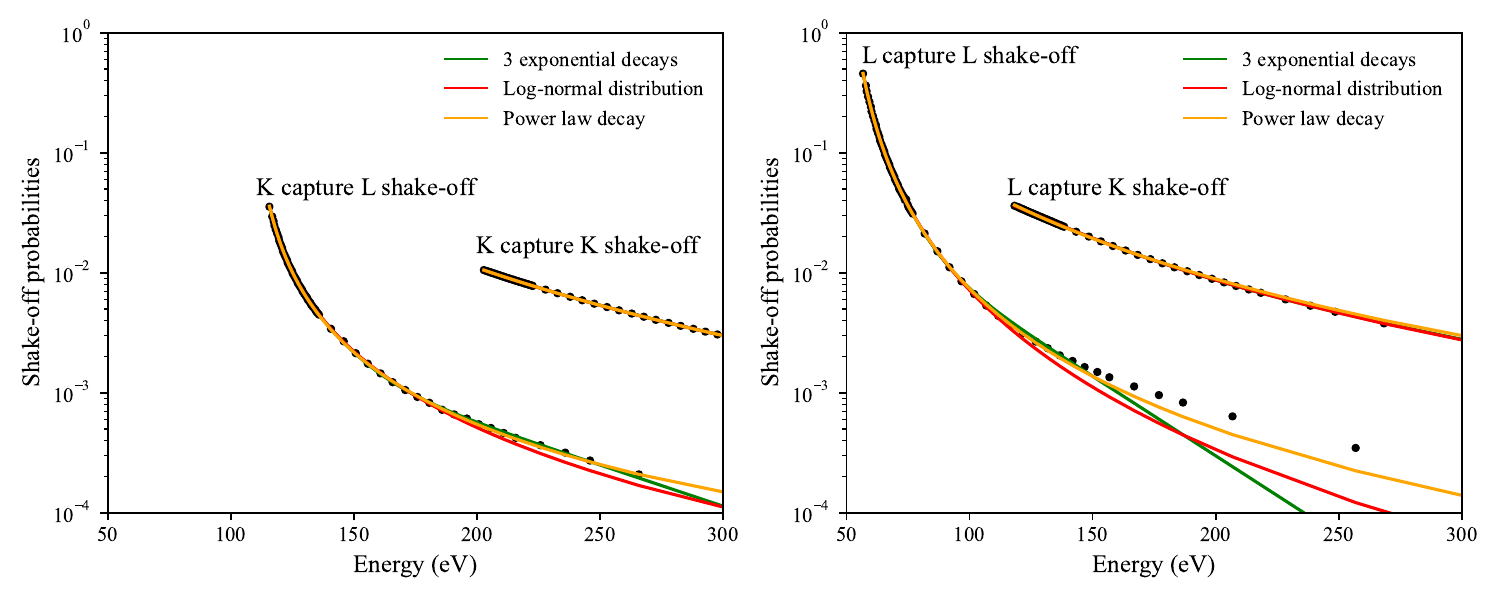}
    \caption{Simulations of the shake-off spectra according to Eq. \ref{shakeup} (dots) after K capture (left panel) and L capture (right panel) compared to analytical approximations by a sum of three exponential functions (green), a log-normal distribution (red) and a power law decay (orange).}
    \label{fig:shakeoff_shapes}
\end{figure*}

For an automated and stable analysis of the phase-III data, the complexity of the fit model should be kept to a minimum. While our \textit{ab initio} SO spectra are computed point by point for different electron energies, in order to keep our model as simple as possible, we have tested several probability density functions to determine whether some of them can reproduce the SO data accurately enough to be used in the final fit. Figure \ref{fig:shakeoff_shapes} shows those analytical functions that could match our calculations (dots) with low reduced chi square. It turns out that a sum of three exponential decays, a log-norm distribution and a power function can all fit the numerical calculations quite well. For the phase-III analysis, we have chosen to describe the K SO spectra with power functions and the L SO spectra with log-normal distributions since they provide the best fit to the data and keep the number of free parameters small \cite{1459}.

\subsection{Double Shake Processes}

The high sensitivity of the BeEST experiment might allow for the observation of subtle effects, and we have, therefore, also calculated the probabilities for double SU and SO. The double SU features are calculated from Eq.~(\ref{shakeup}), taking into account that double SU occurs through an intermediate state and that all possibilities to reach a final state must be considered. For example, an L capture followed by a KL SU to $1s~3s~4s$ can occur in two ways: a transition from one of the two 1s electrons to the 3s orbital with a simultaneous transition from the remaining 2s electron to the 4s orbital, or a transition from the 1s electron to the 4s shell together with a transition from the 2s electron to the 3s shell. We have calculated the probabilities for a wide range of double SU transitions upon electron capture in $^7$Be. Table~\ref{tab:doubleshakeup} shows the most intense ones that are most likely to be observable. However, since L SU energies are small, many of the double SU processes have energies close to those of single-shake processes and are therefore unlikely to be seen as separate peaks in the experiment.
The double shake probabilities were calculated using Eq. \ref{eq:multishake}, either by using $k_e=2$ for KK or LL shakes or by multiplying the single shake probabilities for each individual subshell for the cases of KL SO. The double SO probabilities, as was performed for the single shakes, were computed by subtracting the double SU probabilities (see Table \ref{tab:doubleshakeup}) from the total double shakes (Table \ref{tab:shake}).

\begin{table*}[tb]
\centering
\caption{Double shake-up probabilities after K and L capture. The total is the sum of all double SU channels that were considered in this work. The contribution of SU probabilities from levels not considered in this work is estimated to be negligible.}
\begin{tabular}{ c c c c }
Configuration & L capture KL shake-up & K capture LL shake-up & Energy (eV) \\ 
\hline
$1s 2s 3s$	&	1.54(31)E-04	&		&	118.81(3)	\\
$1s 3s^2$	&	5.17(11)E-06	&	2.47(50)E-06	&	126.96(3)	\\
$1s 3s 4s$	&	2.77(56)E-07	&	1.51(31)E-06	&	128.29(3)	\\
$1s 3s 5s$	&	5.73(12)E-08	&	5.22(11)E-07	&	128.95(3)	\\
$1s 4s^2$	&	2.04(41)E-08	&	2.83(57)E-10	&	131.32(3)	\\
$1s 4s 5s$	&	1.26(26)E-09	&	3.28(66)E-10	&	132.30(3)	\\
$1s 5s^2$	&	3.38(68)E-10	&	3.34(10)E-13	&	133.27(3)	\\
\hline
Total	& 3.11(63)E-04	&	4.5(9)E-06 &	\\
\hline
	&	L capture KK shake-up	&	K capture KL shake-up	&		\\
 \hline
$2s^2 3s$	&	2.51(51)E-06	&	1.46(30)E-04	&	206.77(5)	\\
$2s^2 4s$	&	7.88(16)E-07	&	7.88(16)E-07	&	208.25(5)	\\
$2s 3s^2$	&	1.00(21)E-06	&	2.58(52)E-05	&	217.57(5)	\\
$2s 3s 4s$	&	1.58(32)E-06	&	4.76(96)E-06	&	220.02(6)	\\
$2s 3s 5s$	&	1.33(27)E-08	&	8.18(17)E-07	&	219.82(5)	\\
$2s 4s^2$	&	1.12(23)E-07	&	5.05(11)E-07	&	222.71(6)	\\
$2s 5s^2$	&	2.37(48)E-08	&	5.53(12)E-08	&	224.90(6)	\\
\hline
Total	& 6.0(2)E-06	&	1.79(36)E-04	&		
\end{tabular}
    \label{tab:doubleshakeup}
\end{table*}

\subsection{\textit{Ab initio} simulated spectrum}

\begin{figure*}[tb]
    \centering
    \includegraphics[width=\linewidth]{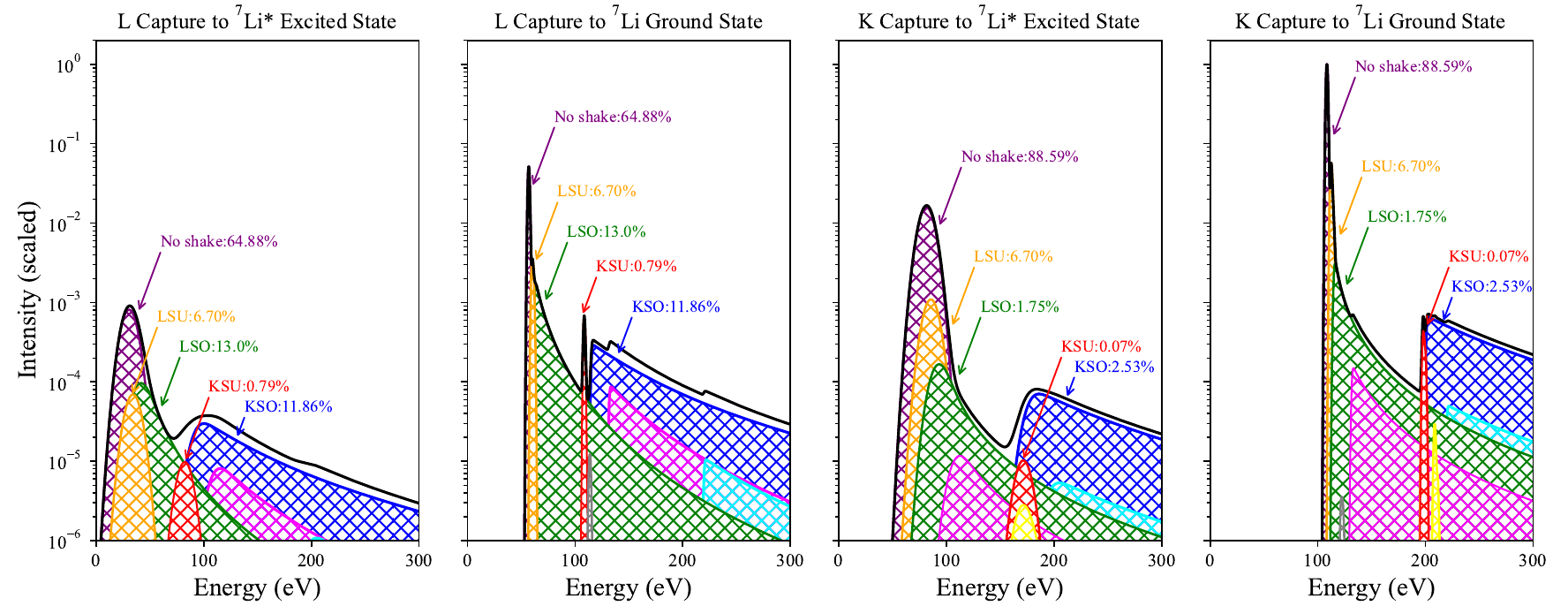}
    \caption{Full \textit{ab initio} simulations of the four $^7$Be decay channels with all SU and shake-off processes included. The two spectra for decay into the $^7$Li ground state are broadened with the detector resolution of 2 eV FWHM, and the decays into the excited state $^7$Li$^*$ are broadened with the measured width of 16.8 eV FWHM to account for the Doppler effect. The double shake processes for K capture are LL shake-off (pink, 0.19\%), KL shake-off (cyan, 0.21\%), LL shake-up (grey, $<$0.01\%) and KL SU (yellow, 0.02\%). Double-shake processes after L capture panels are KL SO (pink, 2.27\%), KK SO (cyan, 0.43\%), KL SU (grey, 0.03\%) and KK SU (yellow, $<$0.01\%).}
    \label{fig:sim_spec_4panels}
\end{figure*}

Figure~\ref{fig:sim_spec_4panels} summarizes the full ab initio calculations of the EC spectra for the four $^7$Be decay channels: K capture to the ground state of $^7$Li (K-GS), K capture to the excited state of $^7$Li$^*$ (K-ES) and the two corresponding L capture channels L-GS and L-ES. It includes all single-electron and two-electron SU and SO processes calculated in the MultiConfiguration Dirac-Fock (MCDF) framework for a set of basis states up to 4s. The spectra are offset by the $^7$Li recoil energy of 56.826(9) eV for decay into the ground state and 28.747(4) eV for decay into the excited state of $^7$Li$^*$ \cite{1345}. The two decay branches into the ground state of $^7$Li are convolved with a Gaussian function with a width of 2 eV FWHM, since the 2 eV energy resolution of the STJ detectors is currently the only fully understood source of broadening in the BeEST experiment. The decay branches into the excited state of $^7$Li$^*$ are convolved with a Gaussian function with a width of 16.84 eV FWHM to account for the Doppler broadening \cite{1459}. The K-GS spectrum is normalized to 1 at its peak value, and the other branches are scaled by the branching ratio of 10.44\% for decay to $^7$Li$^*$ \cite{1431} and the measured L/K capture ratio of 0.07 \cite{1368}. 

The four primary peaks in Fig. \ref{fig:sim_spec_4panels} are shown in dark violet for emphasis, and all SU and SO contributions are colored consistently with L SU in orange, L SO in green, K SU in red and K SO in blue. Under current experimental conditions, double-shake contributions remain at the level of statistical noise and cannot be unambiguously identified, despite appearing in relatively clear regions of the spectrum. The expected relative intensities for each decay channel — represented in gray, pink, yellow, and cyan, as described in the figure captions — contribute less than 2.5\% to the L-capture and less than 0.2\% to the K-capture spectra (Table~\ref{tab:shake} and \ref{tab:doubleshakeup}), and mostly overlap with single-shake features. These components are unlikely to appear as distinct peaks except in very high-statistics spectra, although they may influence the measured intensities of neighboring features. The simulated spectrum shown in Fig.~\ref{fig:sim_spec_4panels} serves as a basis for comparison with the BeEST experimental data.

\section{Discussion} \label{sec:Discussion}

\subsection{Comparison with Experiment}
A direct comparison of the calculated SU and SO spectra with the BeEST experiment is difficult because the experiment does not measure the EC decay of isolated $^7$Be atoms but of $^7$Be implanted into Ta-based STJ sensors. The experimental spectra therefore differ from the calculated ones. Specifically, the measured peaks are significantly wider than the STJ detector resolution of 2 eV FWHM, and peak centroids are consistently lower in energy than expected from the simulations and the literature values. The origin of these differences is currently not fully understood. We speculate that SU of the 5d electrons in the Ta absorber film affects the measured energies and broadens the peaks. Since SU in $^7$Be is stronger after L capture than after K capture, this would affect L capture peaks more strongly than K capture peaks. In addition, STJ sensors may not have the same response to nuclear recoils and to electronic interactions with the same energy, an effect known as "nuclear quenching". This would reduce signals from decays into the $^7$Li ground state differently from those into the $^7$Li$^*$ excited state. For comparison with experiment, we therefore shift the calculated centroids of the peaks and convolve the spectra with Gaussian functions whose width is in accordance to the measured values \cite{1459}. SU and SO spectra are shifted and convolved with Gaussian functions with the same parameters as the corresponding no-shake primary peaks for consistency.

\begin{figure*}[tb]
    \centering
    \includegraphics[width=\linewidth]{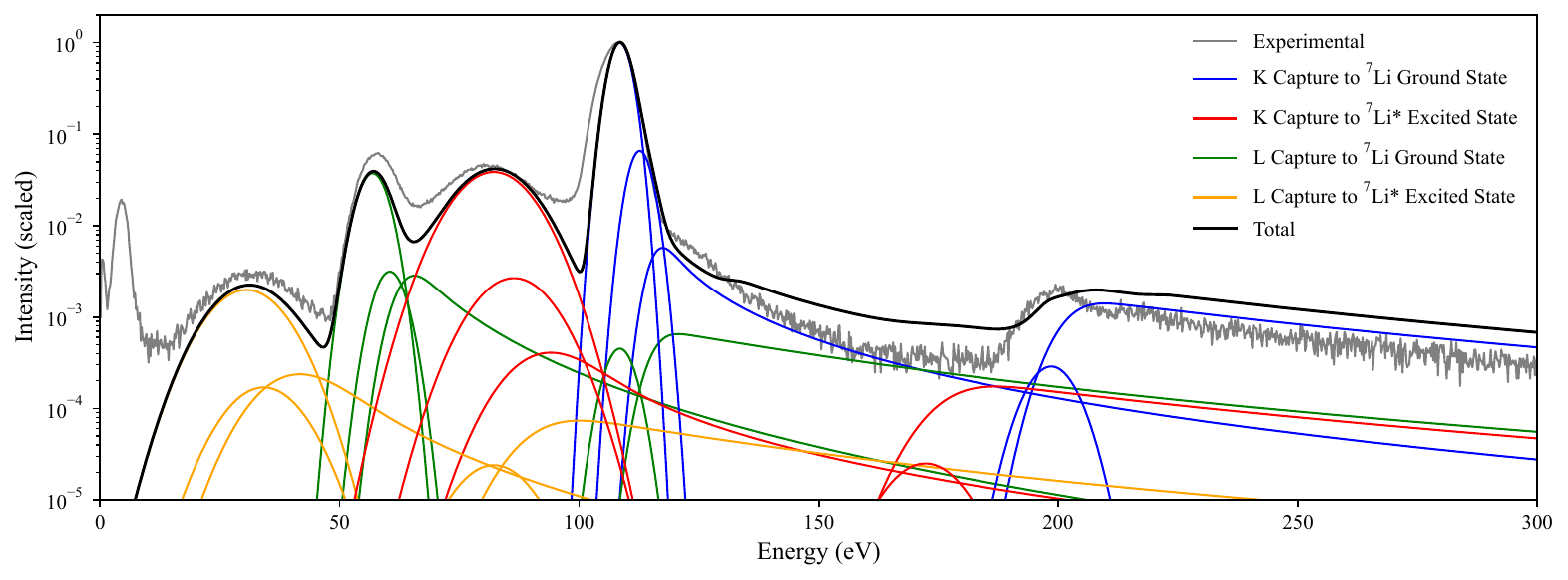}
    \caption{Simulation of the full $^7$Be electron capture spectrum in the 0-300 eV range, normalized to 1 and compared to a normalized one-day spectrum from phase-III of the BeEST experiment. All four primary peaks are accompanied by L shake-up, L shake-off, K shake-up and K shake-off spectra. The atomic structure calculations were shifted in energy and convolved with Gaussian responses to match the experimental data. The L capture spectra were scaled by the L/K capture ratio of 0.070(7) \cite{1431} and the excited-state spectra by the branching ratio of 10.44\% \cite{1368}.}
    \label{fig:sim_spec_5_5eV}
\end{figure*}

Figure \ref{fig:sim_spec_5_5eV} shows the full \textit{ab initio} simulations, shifted and broadened to match the centroids and widths of the experimental data. Both simulations and experimental spectra have been normalized to unity at the K-GS peak at 108.50 eV. The simulation of the L-GS spectrum has been scaled by the L/K ratio of 0.070(7) \cite{1368}, and the two excited-state spectra have been scaled relative to the ground-state spectra by the branching ratio of 10.44\% \cite{1431}. Amplitudes of the SU and SO spectra are not changed from their \textit{ab initio} values, and neither are the shapes of the SO spectra. Figure \ref{fig:sim_spec_5_5eV} illustrates the strengths and limitations of simulating SU and SO in atomic $^7$Be. The shapes of the SO spectra match the observations reasonably well, especially for energies well above the ionization threshold where the influence of the periodic potential of the Ta lattice is small. They also suggest that the energy range from $\sim$160 to $\sim$190 eV is affected by the KSO tail of the L-GS spectrum that had not been included in earlier analyses. The experimental spectrum matches the atomic simulations less well at energies close to the ionization threshold, e.g., in the energy range from $\sim$120 to $\sim$150 eV where the K capture LSO tail dominates. That may not be surprising since transition probabilities in this energy range can be altered significantly based on the chemical environment of the $^7$Li \cite{1463}. This also likely explains why the amplitudes of the SU and SO spectra differ from the measurements, since their normalization is calculated from the overlap between the atomic wavefunctions of $^7$Be and $^7$Li that determine the total shake probability (Sec.~\ref{sec:develop}). This is e.g. visible around $\sim$200 eV where the predicted K capture KSU peak is too small while the K capture KSO tail is too large.

The simulations can, however, shine light on one of the open questions, namely the fact that the fit of the KGS peak requires more than one Voigt component. They suggest that the secondary component in the K-GS peak at higher energy is due to L SU. While the calculated centroid and amplitude of the LSU peak differ slightly from the experiment, the values are of the right order to roughly match the high-energy shoulder of the K-GS peak. The differences are again likely due to solid-state effects. Since the L-GS peak is wider than the K-GS peak, the L capture LSU contribution cannot be resolved as a secondary component at higher energy. On the other hand, we do not yet understand the origin of the third component on the low-energy side of the K-GS peak. We speculate that it might be due to energy loss from lattice damage by the $^7$Li recoil, although other loss mechanisms could also be possible. This low-energy component contains a significant fraction of $\sim$25\% of all events in the K-GS spectrum. Since the areas of the other peaks are scaled relative to the K-GS spectrum, in the absence of the third K-GS component, the computed features contain fewer counts than the experimental data. For the K-ES peak, like in the K-GS peak which constrains its shape, the counts are missing on the low-energy shoulder. 

Finally, the experimental spectra differ from the simulations around $\sim$100 eV. This region is affected by electron escape during the initial relaxation of the Auger electron and depends on the depth distribution of the $^7$Be nuclei relative to the STJ detector surface \cite{1459}. All these effects need to be considered for a full fit of the phase-III spectrum of the BeEST experiment.

\subsection{Fit to Experiment}

For a complete fit to the experimental spectrum, we make two simplifying assumptions. First, we replace the numerical simulations of the SU spectra with a single centroid, because SU is dominated by transitions into the $3s$ orbital and the observed broadening makes the SU spectra indistinguishable from a single Gaussian peak within the accuracy of current measurements (see inset of Fig. \ref{fig:shakeup}). We do not include separate terms for K SU after L capture in the fit, because the final states are identical to L SU after K-capture. Secondly, we replace the ab-initio SO spectra with their analytical approximations (Fig.~\ref{fig:shakeoff_shapes}). The L SO spectra are approximated by log-normal distributions and the K SO spectra by power-law decays with some threshold energy. Amplitudes, decay scales and threshold energies of the SO spectra are allowed to vary as the simplest approximation to account for matrix effects. SO spectra after transitions to the $^7$Li GS and ES are constrained to the same shape and relative intensity, since SO is not expected to depend on the final nuclear state.

In addition, we include a third component to the K-GS peak and constrain the K-ES to the same three components before Doppler broadening. We then add an exponentially modified Gaussian below the two K capture peaks (but not the L capture peaks) to describe the Auger electron escape tail. The escape tails are constrained to the same relative intensity after K-GS and K-ES transitions. The K capture peaks are described by a Voigt function with a natural linewidth of 30 meV to account for the lifetime of the $1s$ hole. This width value is dominated by the Auger rate, and we have computed it by including the Auger transition, the magnetic dipole M1 $1s 2s^2$ $^2 \mathrm{S}_{1/2}$ $\rightarrow 1s^2 2s$ $^2 \mathrm{S}_{1/2}$ as well as two-photon one-electron and two-electron one-photon transitions. Interestingly, the L-GS peak can be fit with a single Gaussian function, most likely because it is somewhat wider than the K-GS peak and its high-energy shoulder overlaps with the broad K-ES peak. We therefore do not include LSU peaks after L capture to avoid overfitting and instead include the LSU contribution in the primary L capture peaks and their SO tails. In addition, the $2s$ hole is sufficiently long-lived to neglect any lifetime broadening for the L capture peaks. Centroids and widths of all peaks are allowed to vary, although the K-ES and L-ES spectra are convolved by the same Gaussian width because Doppler broadening is not expected to depend on the source of the captured electron. SU and SO spectra are convolved with a Gaussian of the same width as their corresponding no-shake peaks. Importantly, we no longer constrain the L/K capture ratio by the literature value of 0.070(7) from phase-II of the BeEST experiment \cite{1368} because we expect the new SO spectra to improve the L/K ratio with reduced systematic uncertainties. 

\begin{figure*}[tb]
    \centering
    \includegraphics[width=\linewidth]{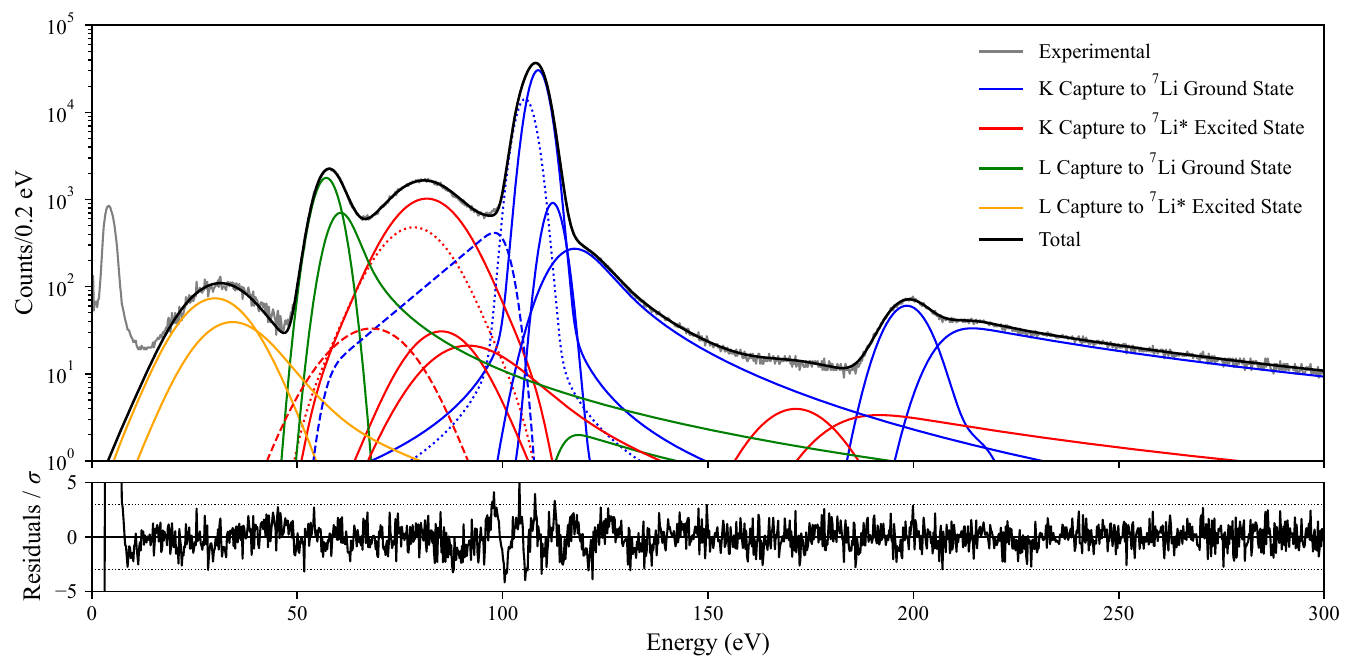}
    \caption{Fit of the shake-up and shake-off spectra (solid lines) to the phase-III data. A third component (dotted) and an electron escape tail (dashed) have to be added to the K-capture peaks to accurately model the spectra. The fit has $\chi^2_{red} \approx$ 1.4 in the fit range between 20--300~eV. The L capture K shake-off curve is not seen due to the fact that it is highly degenerated with the K capture L shake-off curve and becomes quite small in the fit.}
    \label{fig:fit_phase-III}
\end{figure*}

Double-shake events are not observed in the current data set and are therefore not included in the fits. They may be somewhat overestimated in the calculations since their magnitudes are expected to be comparable to the statistical fluctuations. It is possible that these transitions may still be identified once the full phase-III spectrum is unblinded, although it may be difficult to distinguish the KL SO after L-capture from matrix effects, and to distinguish KL SO after K-capture from KGS-KGS pile-up. Details of data processing and spectral fitting are discussed in \cite{1459}.

Note that the BeEST phase-III data have so far only been unblinded outside the region of interest (ROI) for a sterile neutrino search. We therefore scale the unblinded data to the intensity of the spectra in the ROI, so that the statistical accuracy of the data is much higher outside the interval from 20 to 105 eV. 

Under these conditions, the \textit{ab initio} simulations can be fit to the experimental spectrum with reduced chi-squared $\chi^2_{red}\approx 1.4$ (Fig. \ref{fig:fit_phase-III}). The primary reason for having better agreement than directly comparing the simulated spectrum to the data is the adjusted intensity for the various SU and SO contributions. This is not unexpected, especially after L capture, since the \textit{ab initio} spectra are normalized using atomic wavefunctions (Section II. C) that are only approximate inside solids. Most obviously, the K-GS K-SU peak is $\sim$3.4 times stronger and the K-GS K-SO tail is $\sim$50$\%$ weaker than calculated. This makes the K-GS K-SU peak clearly visible at $\sim$200 eV. On the other hand, the measured intensity of 2.04(4)$\%$ for the K-GS L-SU peak is around 3.5 times lower than the computed value of 7.5(5)$\%$, which includes a contribution of 0.79(6)$\%$ due to L-GS K-SU.

\begin{figure}[tb]
    \centering
    \includegraphics[width=\linewidth]{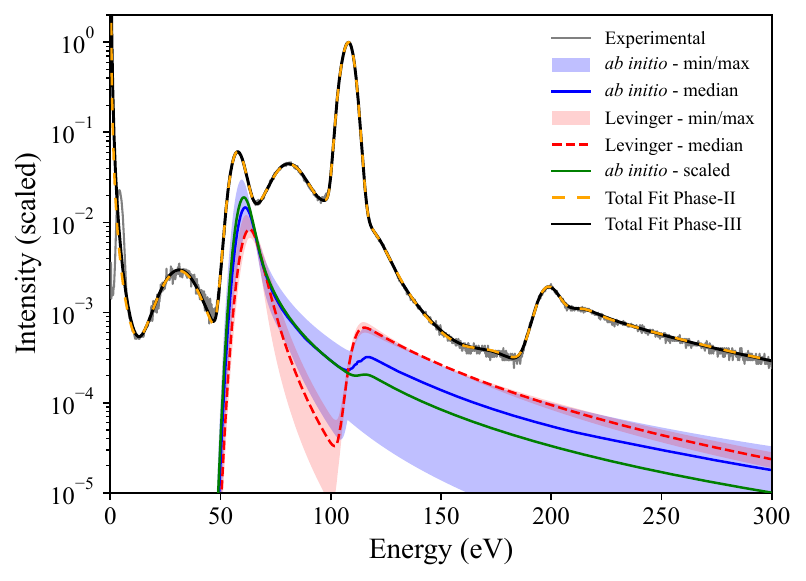}
    \caption{Different analytical approximations of shake-off functions after L capture: Best fit to the unblinded data from phase-III channel 0 (green). Fits to the experimental data with Levinger functions, used in phase-II, are presented in (red). Fits to all 16 channels are presented in (blue) with the shaded area representing the minimum and maximum of the fits to the L capture shake-off peaks. This model dependence currently dominates systematic uncertainties.}
    \label{fig:fits_SO}
\end{figure}

A quantitative comparison of the experimental data with the calculated SO spectra is complicated by the observation that the data can be fit well with a range of parameters (Fig. \ref{fig:fits_SO}). While the K SO spectra dominate above $\sim$200 eV and are therefore constrained by experimental data, L SO spectra overlap with other fit functions that can compensate for different fit parameters. Figure \ref{fig:fits_SO} shows the range of SO functions after L capture that can provide high-quality fits to the data. The best global fit (black) matches the experimental data very well except in [95,115]~eV region inside the main K-GS peak region, and the L capture L SO shapes arising from this fit to all channels are presented in shaded blue representing the maximum, minimum and median (dark blue) of the fits to all 16 channels. Although the component itself is significantly varying, the total fits resulted in equally good $\chi_{red}^2$ values with $\chi_{red}^2$ variations less than 0.03 for each channel. These shapes compare well to ab-initio calculations (green) for energies above the ionization thresholds, as seen by the matching slopes. In this energy range, matrix effects should be small, while the chemical environment is known to affect transition probabilities strongly closer to the ionization threshold \cite{1463}. Also, the sudden approximation is expected to provide better results for higher energies of the ejected electron. The blue curves show that alternative parametrization of the SO curves also provide good fits to the data, and even the earlier non-relativistic function (Levinger) without electron correlations can be used to reproduce the data well (red).
We found that adjusting the decay scale of the fit function serves as an acceptable approximation for describing the influence of matrix effects in the SO probabilities. In addition, both no-shake L-capture and the onset of the L-SO tail contribute to the L-GS peak at $\sim$57 eV, and their relative contribution are only weakly constrained by the experimental data (Fig. \ref{fig:fits_SO}). This contributes a systematic uncertainty of $\pm 4.9\%$ to the total L SO probability. We can fit the >120~eV region with $\chi_{red}^2\approx 1$ for L-GS LSO+SU probabilities between 20 and 40$\%$, compared to a calculated value of 18.7(9)$\%$ for the total single shake, since both SU and SO are included in the L-GS LSO+SU fit. A comparison of the probabilities of all shake processes from the phase-III fits and the atomic structure calculations of this work is shown in Table \ref{tab:shake_probs}. 

\begin{table}[h]
\centering
\caption{Comparison of the calculated and experimental values of the total probabilities for single shake-up and shake-off channels after K and L electron capture in $^7$Be. The number inside brackets represent the absolute statistical uncertainty at the last significant digit.}
\begin{tabular}{ l c c}
Process & \multicolumn{2}{c}{Shake probabilities} \\ 
 & Theory & Experiment \\ 
\hline
K capture K Shake-up &	$ 0.00066(5) $ & $ 0.00278(3) $ \\
K capture L Shake-up &	$ 0.067(5) $ & $0.0204(4) $ \\
L capture K Shake-up &	$ 0.0079(6) $ & $-$ \\
L capture L Shake-up &	$ 0.067(5) $ & $-$ \\
\hline
K capture K Shake-off &	$ 0.0251(10) $ & $0.0118(8) $ \\
K capture L Shake-off &	$ 0.016(8) $ & $0.0197(4) $ \\
L capture K Shake-off &	$ 0.114(4) $ & $0.0203(4) $ \\
L capture L Shake-off &	$ 0.120(11) $ & $0.314(3)^a $ \\
\end{tabular}
\\
\footnotetext[1]{The experimental L capture L Shake-off should be compared to the theoretical total single L capture L shake probability of $0.187(9) \%$.}
    \label{tab:shake_probs}
\end{table}


One inconsistency is that current energy-unconstrained fits predict a L-GS K-SO threshold 12.7(6) eV above the main K-GS peak, while the K-GS L-SO onset sits at only 0.17 eV above the main K-GS peak. This is unphysical, since both channels have the same intermediate state, $1s 2s$, hence they should have the same energy threshold. This likely reflects a breakdown of the model close to the ionization threshold where the influence of the Ta matrix is felt most strongly. We have thus constrained both the L-GS K-SO and K-GS L-SO thresholds to be the same. Due to the high overlap of these two curves, together with the L-GS L-SO and Auger escape tails, the fit results were highly correlated. A Monte-Carlo simulation of the Auger escape tail allowed us to constrain the fit parameters resulting in the fit values presented in Fig. \ref{fig:fit_phase-III}. 
Future work on matrix effects will determine which combination of fit values describes the $^7$Be-$^7$Li-Ta system most accurately.

Still, the fit quality demonstrates that the analytical functions are viable approximations for the \textit{ab initio} SO spectra within the accuracy of this measurement, although their parameters have to be altered slightly from the \textit{ab initio} values to account for matrix effects. The fits show that K-SO after L capture cannot be neglected and contributes to the experimental spectrum between ~160 and ~190 eV. Similarly, as long as the spectra are broadened well beyond the detector resolution of ~2 eV FWHM, SU spectra can be approximated by a single Gaussian function of the appropriate width. This identifies the high-energy component of the K-GS peak as being due to L SU and solves one of the open questions in the BeEST experiment.

\subsection{The L/K Capture Ratio}

The SU and SO calculations enable an improved assessment of the L/K capture ratio for the decay of $^7$Be in Ta. This ratio is important because it enters calculations of $^7$Be decay in stellar environments that are the primary source of cosmological $^7$Li production \cite{1445}. Our earlier analysis based on the phase-II spectrum of the BeEST experiment \cite{1368} was limited by systematic uncertainties in the SO spectra and the background due to gamma interactions in the Si substrate beneath the STJ detectors. This substrate background has been removed by coincidence vetoing in phase-III \cite{1459}. In addition, the calculations of the SO spectra now include relativistic and many-electron effects that had been left out earlier.

Although the shake spectra have so far only been calculated for isolated atoms, they identify several errors in our previous analysis. First, the calculations show that shake effects are stronger after L capture than after K capture (Tables \ref{tab:shakeup} and \ref{tab:doubleshakeup}), in agreement with earlier calculations \cite{1408, 1440}. Fits that constrained the SO spectra after L and K capture to the same relative intensity, which produce a lower L/K ratio, should therefore not have been included in the previous analysis. Secondly, the L SO spectra extend to higher energies than predicted by the Levinger function we used in the earlier analysis (Fig. \ref{fig:shakeoff_comp}). Since shake effects are stronger after L capture, this also leads to a higher L/K ratio since a larger fraction of events between ~60 and ~100 eV is now attributed to L-SO (Fig. \ref{fig:fits_SO}). Thirdly, the K SO spectrum after L capture was not included in earlier calculations \cite{1421}, and we had therefore assumed it to be negligible. Our new calculations show that this assumption was not justified (Fig. \ref{fig:shakeoff_comp}). Finally, phase-II used only a single STJ pixel and therefore could not reject substrate events by coincidence vetoing. This led us to erroneously attribute the spectral background above ~160 eV to gamma interactions in the Si substrate \cite{1368}. After anti-coincidence vetoing all substrate events in the phase-III data, it is apparent that this spectral region is in fact dominated by SU and SO events.

\begin{figure}[tb]
    \centering
    \includegraphics[width=\linewidth]{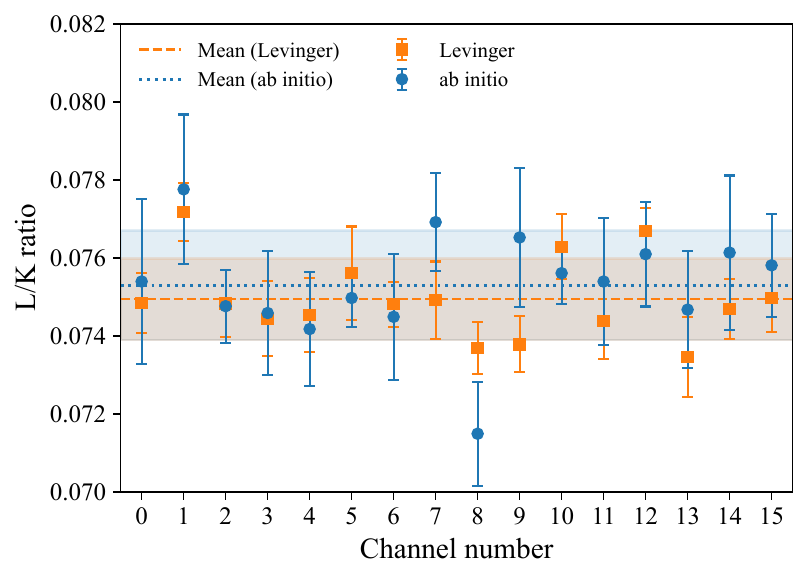}
    \caption{L/K capture ratio for different STJ detector pixels in phase-III of the BeEST experiment. The \textit{ab initio} calculations (blue) produce a slightly higher value of the L/K ratio than the Levinger functions (orange) used in the past \cite{1368}.}
    \label{fig:LKratio}
\end{figure}

With the gamma background removed, the analytical approximations to the new SO spectra produce the best fit of the spectrum in Figure \ref{fig:fit_phase-III} for an L/K electron capture ratio of $0.0756 \pm 0.0020$. 
Surprisingly, this value is consistent with the mean value of the L/K ratio for the phase-III unblinded data using the Levinger functions as was done in phase-II (Figure \ref{fig:LKratio}). It is larger than our earlier value of $0.070 \pm 0.007$ primarily because of the improved SU spectra and the inclusion of new shake features in the fit, although it is contained in the error bar of our previous measurement. This shows that, due to the complexity of the spectra in the region between 50 and 120 eV, a good fit can be attained even if the models of the SO distributions are quite distinct.
 The systematic uncertainty of the new L/K ratio is thus determined by the choice of the fit functions and the range of their parameters that produce plausible fits to the data. This range turns out to be not insignificant, since SO probabilities are small and much of the SO spectra are hidden under the primary peaks (Fig. \ref{fig:fits_SO}). 
 Among the different fits, we consider those most plausible that match the shape of the \textit{ab initio} spectrum for energies well above the SO threshold, where matrix effects are expected to be small. These fits are used in \cite{1459}. They show a different SO probability near the ionization threshold compared to the \textit{ab initio} calculations based on the free electron approximation. This appears plausible as it is also seen in X-ray absorption spectra \cite{1463} and the use of the sudden approximation hinders the possibility of having correct near-threshold shake probabilities. We cannot exclude the possibility that matrix effects will further alter the SO spectra, although we expect the impact on the L/K ratio to be within the current uncertainties. Therefore, we now recommend a new value of $0.0756 \pm 0.0020$ for the L/K capture ratio of $^7$Be in Ta.

\subsection{Outlook}

Current calculations of SU and SO spectra in electron capture (EC) decay, including those presented in this work, are limited to isolated atoms \cite{1370,1419,1445}. Although this approach allows for highly accurate computations of many-electron wavefunctions, it does not include matrix effects from the detector material, which inevitably alter wavefunctions and associated atomic parameters such as energy levels, transition rates, and SU and SO probabilities. Incorporating these matrix effects into future calculations is therefore a crucial next step. Site-specific properties of the Be and Li wavefunctions and orbital hybridization with the detector matrix will likely be needed to achieve the precision required for a theoretically constrained background model of the BeEST spectrum.

In recent years, the BeEST collaboration has conducted density functional theory (DFT) calculations to refine electron binding energies \cite{1408} and nuclear electron capture rates and shaking processes \cite{1440}. These results have already strengthened the phase-III data analysis, and further developments will be instrumental for phase-IV. The computation of shake probabilities beyond the sudden approximation regime may also prove important, especially near SO thresholds where the electron escapes on a time scale comparable to the electron relaxation time in the lattice. We plan to examine low-energy ($\lesssim$10–20 eV) continuum electrons using a time-dependent or continuum-distorted-wave approach, or alternatively a time dependent density functional theory embedding framework, to constrain potential sudden approximation biases near the ionization thresholds.

Similar modeling will likely be important for precise EC measurements in other high-resolution detectors, including those used for neutrino mass experiments based on the EC decay of $^{163}$Ho implanted into cryogenic Au \cite{1424} or Sn \cite{1376} detectors. Although these experiments primarily aim to measure the endpoint spectrum, SO effects can modify the spectral shape in this region. Achieving sub-eV uncertainties, as targeted by these collaborations, demands a rigorous understanding of SO probabilities in different materials. 

\section{Summary} \label{sec:summary}

We have calculated the first full spectrum of the EC decay of atomic $^7$Be using correlated wavefunctions up to the 4s orbital. The calculations include all single SU, SO and double-shake excitations. The calculations show that shaking probabilities after L capture are much higher than their K capture counterparts and contribute to around 30\% of all L capture events. Specifically, K SO after L capture cannot be neglected in the EC spectra.
In the BeEST experiment, SU transitions appear as high-energy components of the K and L capture peaks and can be modeled by broadened Gaussian or Voigt functions. SO spectra can be approximated by power law and log-normal distributions to reduce the number of free parameters in the fits and thus the systematic uncertainties in the search for physics beyond the standard model. No spectral features are predicted that could mimic a sterile neutrino signal in the energy range of interest between $\sim$60 and $\sim$108 eV. Double SU probabilities are currently too low to be seen, even in high-statistics spectra, and, although double SO contributions remain below the current detection threshold, their inclusion in the simulations produces relative spectral differences of approximately 24\% around 136 eV and 12\% near 225 eV. This indicates that, with the improved statistics expected in phase-IV data, such structures could become experimentally accessible.

So far, these calculations are limited to wavefunctions of isolated atoms. This likely accounts for the discrepancies between the calculated and measured SU and SO probabilities, since the wavefunctions are altered by matrix effects from the detector materials into which the $^7$Be is implanted. This likely also affects details of the spectral shape, especially near the ionization threshold, where matrix effects are expected to be stronger. Incorporating these effects into the calculations will be the focus of future work. Matrix effects might also help explain the unexpected peak broadening as SU of Ta 5d electrons into unoccupied empty states, similar to the low-energy L SU transitions that have been found to be stronger in atomic $^7$Be.
Still, the calculations allow a re-assessment of the L/K capture ratio in $^7$Be, and we now recommend a value of 0.0756(20) for this ratio in tantalum.
\section*{Acknowledgements} \label{sec:acknowledgements}
This work has been financially supported by Fundação para a Ciência e Tecnologia (FCT) (Portugal) under research center grants UIDB/04559/2025 (LIBPhys) and LA/P/0117/2020 (Associated Laboratory LA-REAL). J. M and J.P.S acknowledge the support of EMPIR, Germany, under Contract No. 20FUN04 PrimA-LTD. The EMPIR initiative is co-funded by the European Union’s Horizon 2020 research and innovation programme and the EMPIR, Germany participating States. Part of this work has been carried out under the High Performance Computing Chair - a R\&D infrastructure (based at the University of Évora; PI: M. Avillez), endorsed by Hewlett Packard Enterprise (HPE), and involving a consortium of higher education institutions (University of Algarve, University of Évora, NOVA University Lisbon, and University of Porto), research centres (CIAC, CIDEHUS, CHRC), enterprises (HPE, ANIET, ASSIMAGRA, Cluster Portugal Mineral Resources, DECSIS, FastCompChem, GeoSense, GEOtek, Health Tech, Starkdata), and public/private organizations (Alentejo Tourism-ERT, KIPT Colab).
 The BeEST experiment is funded in part by the Gordon and Betty Moore Foundation (10.37807/GBMF11571), the DOE-SC Office of Nuclear Physics under Award Numbers DE-SC0021245 and DE-FG02-93ER40789, and the LLNL Laboratory Directed Research and Development program through Grants No. 19-FS-027 and No. 20-LW-006. TRIUMF receives federal funding via a contribution agreement with the National Research Council of Canada. The theoretical work was performed as part of the European Metrology Programme for Innovation and Research (EMPIR) Projects No. 17FUN02 MetroMMC and No. 20FUN09 PrimA-LTD. This work was performed under the auspices of the U.S. Department of Energy by Lawrence Livermore National Laboratory under Contract No. DE-AC52- 07NA27344.

\bibliography{MGuerra}

@Article{163,
   Author = {K. Aimi and T. Fujiwara and S. Ando},
   Title = {A conformational study of aromatic imide compounds. part 1. Compounds containing diphenyl ether and benzophenone moieties},
   Journal = {Journal of Molecular Structure},
   Volume = {602-603},
   Number = {0},
   Pages = {405-416},
   Year = {2002} }

@Article{1440,
   Author = {A. Andoche and L. Mouawad and P. A. Hervieux and X. Mougeot and J. Machado and J. P. Santos},
   Title = {Influence of atomic modeling on electron capture and shaking processes},
   Journal = {Physical Review A},
   Volume = {109},
   Number = {3},
   Pages = {032826},
   Year = {2024} }

@Article{1414,
   Author = {I. Angeli},
   Title = {A consistent set of nuclear rms charge radii: properties of the radius surface R(N,Z)},
   Journal = {Atomic Data and Nuclear Data Tables},
   Volume = {87},
   Number = {2},
   Pages = {185-206},
   Year = {2004} }

@Article{1439,
   Author = {I. Angeli and K. P. Marinova},
   Title = {Table of experimental nuclear ground state charge radii: An update},
   Journal = {Atomic Data and Nuclear Data Tables},
   Volume = {99},
   Number = {1},
   Pages = {69-95},
   Year = {2013} }

@Article{1413,
   Author = {G. Audi and A. H. Wapstra and C. Thibault},
   Title = {The Ame2003 atomic mass evaluation: (II). Tables, graphs and references},
   Journal = {Nuclear Physics A},
   Volume = {729},
   Number = {1},
   Pages = {337-676},
   Year = {2003} }

@Book{1441,
   Author = {V. Barger and D. Marfatia and K. Whisnant},
   Title = {The Physics of Neutrinos},
   Publisher = {Princeton University Press},
      Year = {2012} }

@Article{1453,
   Author = {R. Bhandari and G. Bollen and T. Brunner and N. D. Gamage and A. Hamaker and Z. Hockenbery and M. H. Gamage and D. K. Keblbeck and K. G. Leach and D. Puentes and M. Redshaw and R. Ringle and S. Schwarz and C. S. Sumithrarachchi and I. Yandow},
   Title = {First direct $^{7}\mathrm{Be}$ electron-capture $Q$-value measurement toward high-precision searches for neutrino physics beyond the Standard Model},
   Journal = {Physical Review C},
   Volume = {109},
   Number = {2},
   Pages = {L022501},
   Year = {2024} }

@Article{1463,
   Author = {A. Braun and H. Wang and J. Shim and S. S. Lee and E. J. Cairns},
   Title = {Lithium K(1s) synchrotron NEXAFS spectra of lithium-ion battery cathode, anode and electrolyte materials},
   Journal = {Journal of Power Sources},
   Volume = {170},
   Number = {1},
   Pages = {173-178},
   Year = {2007} }

@Article{1311,
   Author = {T. A. Carlson and C. W. Nestor},
   Title = {Calculation of Electron Shake-Off Probabilities as the Result of X-Ray Photoionization of the Rare Gases},
   Journal = {Physical Review A},
   Volume = {8},
   Number = {6},
   Pages = {2887-2894},
   Year = {1973} }

@Article{1430,
   Author = {T. A. Carlson and C. W. Nestor and T. C. Tucker and F. B. Malik},
   Title = {Calculation of Electron Shake-Off for Elements from $Z=2$ to 92 with the Use of Self-Consistent-Field Wave Functions},
   Journal = {Physical Review},
   Volume = {169},
   Number = {1},
   Pages = {27-36},
   Year = {1968} }

@Article{38,
   Author = {E. Clementi and D. L. Raimondi and W. P. Reinhard},
   Title = {ATOMIC SCREENING CONSTANTS FROM SCF FUNCTIONS .2. ATOMS WITH 37 TO 86 ELECTRONS},
   Journal = {Journal of Chemical Physics},
   Volume = {47},
   Number = {4},
   Pages = {1300-&},
   Year = {1967} }

@Article{1444,
   Author = {B. Crasemann and M. H. Chen and J. P. Briand and P. Chevallier and A. Chetioui and M. Tavernier},
   Title = {Atomic electron excitation probabilities during orbital electron capture by the nucleus},
   Journal = {Physical Review C},
   Volume = {19},
   Number = {3},
   Pages = {1042-1046},
   Year = {1979} }

@Article{1445,
   Author = {P. Das and A. Ray},
   Title = {Terrestrial $^{7}\mathrm{Be}$ decay rate and $^{8}\mathrm{B}$ solar neutrino flux},
   Journal = {Physical Review C},
   Volume = {71},
   Number = {2},
   Pages = {025801},
   Year = {2005} }

@Article{92,
   Author = {J. P. Desclaux},
   Title = {{A multiconfiguration relativistic DIRAC-FOCK program}},
   Journal = {Computer Physics Communications},
   Volume = {9},
   Number = {1},
   Pages = {31-45},
   Year = {1975} }

@Article{1415,
   Author = {J. A. Detwiler and R. G. H. Robertson},
   Title = {Shake-up and shake-off effects in neutrinoless double-$\ensuremath{\beta}$ decay},
   Journal = {Physical Review C},
   Volume = {107},
   Number = {4},
   Pages = {L042501},
   Year = {2023} }

@Article{1427,
   Author = {S. Dodelson and L. M. Widrow},
   Title = {Sterile neutrinos as dark matter},
   Journal = {Physical Review Letters},
   Volume = {72},
   Number = {1},
   Pages = {17-20},
   Year = {1994} }

@Article{1447,
   Author = {A. Faessler and L. Gastaldo and F. Šimkovic},
   Title = {Neutrino mass, electron capture, and the shake-off contributions},
   Journal = {Physical Review C},
   Volume = {95},
   Number = {4},
   Pages = {045502},
   Year = {2017} }

@Article{1376,
   Author = {M. Faverzani and B. Alpert and D. Backer and D. Bennet and M. Biasotti and C. Brofferio and V. Ceriale and G. Ceruti and D. Corsini and P. K. Day and M. De Gerone and R. Dressler and E. Ferri and J. Fowler and E. Fumagalli and J. Gard and F. Gatti and A. Giachero and J. Hays-Wehle and S. Heinitz and G. Hilton and U. Köster and M. Lusignoli and M. Maino and J. Mates and S. Nisi and R. Nizzolo and A. Nucciotti and A. Orlando and L. Parodi and G. Pessina and G. Pizzigoni and A. Puiu and S. Ragazzi and C. Reintsema and M. Ribeiro-Gomez and D. Schmidt and D. Schuman and F. Siccardi and M. Sisti and D. Swetz and F. Terranova and J. Ullom and L. Vale},
   Title = {The HOLMES Experiment},
   Journal = {Journal of Low Temperature Physics},
   Volume = {184},
   Number = {3},
   Pages = {922-929},
   Year = {2016} }

@Article{1368,
   Author = {S. Fretwell and K. G. Leach and C. Bray and G. B. Kim and J. Dilling and A. Lennarz and X. Mougeot and F. Ponce and C. Ruiz and J. Stackhouse and S. Friedrich},
   Title = {Direct Measurement of the $^{7}\mathrm{Be}$ L/K Capture Ratio in Ta-Based Superconducting Tunnel Junctions},
   Journal = {Physical Review Letters},
   Volume = {125},
   Number = {3},
   Pages = {032701},
   Year = {2020} }

@Article{1345,
   Author = {S. Friedrich and G. B. Kim and C. Bray and R. Cantor and J. Dilling and S. Fretwell and J. A. Hall and A. Lennarz and V. Lordi and P. Machule and D. McKeen and X. Mougeot and F. Ponce and C. Ruiz and A. Samanta and W. K. Warburton and K. G. Leach},
   Title = {Limits on the Existence of sub-MeV Sterile Neutrinos from the Decay of $^{7}\mathrm{Be}$ in Superconducting Quantum Sensors},
   Journal = {Physical Review Letters},
   Volume = {126},
   Number = {2},
   Pages = {021803},
   Year = {2021} }

@Article{1424,
   Author = {L. Gastaldo and K. Blaum and K. Chrysalidis and T. Day Goodacre and A. Domula and M. Door and H. Dorrer and C. E. Düllmann and K. Eberhardt and S. Eliseev and C. Enss and A. Faessler and P. Filianin and A. Fleischmann and D. Fonnesu and L. Gamer and R. Haas and C. Hassel and D. Hengstler and J. Jochum and K. Johnston and U. Kebschull and S. Kempf and T. Kieck and U. Köster and S. Lahiri and M. Maiti and F. Mantegazzini and B. Marsh and P. Neroutsos and Y. N. Novikov and P. C. O. Ranitzsch and S. Rothe and A. Rischka and A. Saenz and O. Sander and F. Schneider and S. Scholl and R. X. Schüssler and C. Schweiger and F. Simkovic and T. Stora and Z. Szücs and A. Türler and M. Veinhard and M. Weber and M. Wegner and K. Wendt and K. Zuber},
   Title = {The electron capture in 163Ho experiment – ECHo},
   Journal = {The European Physical Journal Special Topics},
   Volume = {226},
   Number = {8},
   Pages = {1623-1694},
   Year = {2017} }

@Article{1,
   Author = {G. Glupe and W. Mehlhorn},
   Title = {Absolute Electron Impact Ionization cross Sections of N, O and Ne},
   Journal = {Journal de Physique},
   Volume = {C4},
   Number = {32},
   Pages = {40},
   Year = {1971} }

@Article{1153,
   Author = {M. Guerra and P. Amaro and J. P. Santos and P. Indelicato},
   Title = {{Relativistic calculations of screening parameters and atomic radii of neutral atoms}},
   Journal = {Atomic Data and Nuclear Data Tables},
   Year = {2017} }

@Article{2,
   Author = {D. H. H. Hoffmann and C. Brendel and H. Genz and W. Low and S. Muller and A. Ritcher},
   Journal = {Zeitschrift fuer Physik A: Atoms and Nuclei},
   Volume = {293},
   Pages = {187},
   Year = {1979} }

@Article{1435,
   Author = {I. Hornyák and L. Adamowicz and S. Bubin},
   Title = {Ground and excited $^{1}S$ states of the beryllium atom},
   Journal = {Physical Review A},
   Volume = {100},
   Number = {3},
   Pages = {032504},
   Year = {2019} }

@Article{512,
   Author = {P. Indelicato},
   Title = {Nonperturbative evaluation of some QED contributions to the muonic hydrogen n=2 Lamb shift and hyperfine structure},
   Journal = {Physical Review A},
   Volume = {87},
   Number = {2},
   Pages = {022501},
   Year = {2013} }

@Article{1388,
   Author = {P. Indelicato},
   Title = {QED tests with highly charged ions},
   Journal = {Journal of Physics B: Atomic, Molecular and Optical Physics},
   Volume = {52},
   Number = {23},
   Pages = {232001},
   Year = {2019} }

@misc{1465,
   Author = {P. Indelicato},
   Title = {The MDFGME Multiconfiguration Dirac-Fock code},
   URL= {https://www.lkb.fr/exoticions/research/research-topics/mdfgme/} }

@Article{93,
   Author = {P. Indelicato and J. P. Desclaux},
   Title = {{Multiconfiguration Dirac-Fock calculations of transition energies with QED corrections in three-electron ions}},
   Journal = {Physical Review A},
   Volume = {42},
   Number = {9},
   Pages = {5139},
   Year = {1990} }

@Article{1417,
   Author = {F. F. Karpeshin and M. B. Trzhaskovskaya},
   Title = {Shake-off in the $^{164}\mathrm{Er}$ neutrinoless double-electron capture and the dark matter puzzle},
   Journal = {Physical Review C},
   Volume = {107},
   Number = {4},
   Pages = {045502},
   Year = {2023} }

@Article{1370,
   Author = {A. S. Kheifets and B. Igor},
   Title = {Double shake-off model for the triple photoionization of beryllium},
   Journal = {Journal of Physics B: Atomic, Molecular and Optical Physics},
   Volume = {36},
   Number = {13},
   Pages = {L211},
   Year = {2003} }

@Article{1459,
   Author = {I. Kim and C. Bray and A. Marino and C. Stone-Whitehead and A. Lamm and R. Abells and P. Amaro and A. Andoche and R. Cantor and D. Diercks and S. Fretwell and A. Gillespie and M. Guerra and A. Hall and C. N. Harris and J. T. Harris and C. Hinkle and L. M. Hayen and P.-A. Hervieux and G.-B. Kim and K. G. Leach and A. Lennarz and V. Lordi and J. Machado and D. McKeen and X. Mougeot and F. Ponce and C. Ruiz and A. Samanta and J. P. Santos and J. Smolsky and J. Taylor and J. Templet and S. Upadhyayula and L. Wagner and W. K. Warburton and B. Waters and S. Friedrich},
   Title = {Signal processing and spectral modeling for the BeEST experiment},
   Journal = {Physical Review D},
   Volume = {111},
   Number = {5},
   Pages = {052010},
   Year = {2025} }

@Article{1419,
   Author = {A. G. Kochur and V. A. Popov},
   Title = {Probabilities of multiple shake processes in sudden approximation},
   Journal = {Journal of Physics B: Atomic, Molecular and Optical Physics},
   Volume = {39},
   Number = {16},
   Pages = {3335},
   Year = {2006} }

@Article{1462,
   Author = {A. G. Kochur and V. A. Popov},
   Title = {The relative role of shake-up and shake-off processes in additional monopole excitation of L and M electrons due to inner atomic shell ionization},
   Journal = {Optics and Spectroscopy},
   Volume = {100},
   Number = {5},
   Pages = {645-651},
   Year = {2006} }

@Article{1421,
   Author = {J. S. Levinger},
   Title = {Effects of Radioactive Disintegrations on Inner Electrons of the Atom},
   Journal = {Physical Review},
   Volume = {90},
   Number = {1},
   Pages = {11-25},
   Year = {1953} }

@Article{1466,
   Author = {R. Loetzsch and H. F. Beyer and L. Duval and U. Spillmann and D. Banaś and P. Dergham and F. M. Kröger and J. Glorius and R. E. Grisenti and M. Guerra and A. Gumberidze and R. Heß and P. M. Hillenbrand and P. Indelicato and P. Jagodzinski and E. Lamour and B. Lorentz and S. Litvinov and Y. A. Litvinov and J. Machado and N. Paul and G. G. Paulus and N. Petridis and J. P. Santos and M. Scheidel and R. S. Sidhu and M. Steck and S. Steydli and K. Szary and S. Trotsenko and I. Uschmann and G. Weber and T. Stöhlker and M. Trassinelli},
   Title = {Testing quantum electrodynamics in extreme fields using helium-like uranium},
   Journal = {Nature},
   Volume = {625},
   Number = {7996},
   Pages = {673-678},
   Year = {2024} }

@Article{1409,
   Author = {P. J. Mohr},
   Title = {Self-energy correction to one-electron energy levels in a strong Coulomb field},
   Journal = {Physical Review A},
   Volume = {46},
   Number = {7},
   Pages = {4421-4424},
   Year = {1992} }

@Article{1410,
   Author = {P. J. Mohr and Y.-K. Kim},
   Title = {Self-energy of excited states in a strong Coulomb field},
   Journal = {Physical Review A},
   Volume = {45},
   Number = {5},
   Pages = {2727-2735},
   Year = {1992} }

@Article{1411,
   Author = {P. J. Mohr and G. Soff},
   Title = {Nuclear size correction to the electron self-energy},
   Journal = {Physical Review Letters},
   Volume = {70},
   Number = {2},
   Pages = {158-161},
   Year = {1993} }

@Article{1363,
   Author = {T. V. B. Nguyen and H. A. Melia and F. I. Janssens and C. T. Chantler},
   Title = {Multiconfiguration Dirac-Hartree-Fock theory for copper $K\ensuremath{\alpha}$ and $K\ensuremath{\beta}$ diagram lines, satellite spectra, and ab initio determination of single and double shake probabilities},
   Journal = {Physical Review A},
   Volume = {105},
   Number = {2},
   Pages = {022811},
   Year = {2022} }

@Article{1437,
   Author = {M. Przybytek and M. Lesiuk},
   Title = {Correlation energies for many-electron atoms with explicitly correlated Slater functions},
   Journal = {Physical Review A},
   Volume = {98},
   Number = {6},
   Pages = {062507},
   Year = {2018} }

@Article{1030,
   Author = {P. Pyykko and M. Atsumi},
   Title = {{Molecular Single-Bond Covalent Radii for Elements 1-118}},
   Journal = {Chemistry – A European Journal},
   Volume = {15},
   Number = {1},
   Pages = {186-197},
   Year = {2009} }

@Article{1422,
   Author = {R. G. H. Robertson and V. Venkatapathy},
   Title = {Shakeup and shakeoff satellite structure in the electron spectrum of $^{83}\mathrm{Kr}^{m}$},
   Journal = {Physical Review C},
   Volume = {102},
   Number = {3},
   Pages = {035502},
   Year = {2020} }

@Article{1408,
   Author = {A. Samanta and S. Friedrich and K. G. Leach and V. Lordi},
   Title = {Material Effects on Electron-Capture Decay in Cryogenic Sensors},
   Journal = {Physical Review Applied},
   Volume = {19},
   Number = {1},
   Pages = {014032},
   Year = {2023} }

@Article{1452,
   Author = {C. Santonastaso and N. Casali and L. Di Benedetto and V. Boldrini and R. Buompane and M. Canino and V. Carrano and A. Formicola and L. Gialanella and M. Laubenstein and H. C. Neitzert and M. Pieruccini and G. Porzio and A. Rubino},
   Title = {Precise measurement of the 7Be electron capture decay half-life in silicon carbide},
   Journal = {Journal of Physics G: Nuclear and Particle Physics},
   Volume = {52},
   Number = {3},
   Pages = {035101},
   Year = {2025} }

@Article{1418,
   Author = {V. A. Sevestrean and O. Niţescu and S. Ghinescu and S. Stoica},
   Title = {Self-consistent calculations for atomic electron capture},
   Journal = {Physical Review A},
   Volume = {108},
   Number = {1},
   Pages = {012810},
   Year = {2023} }

@incollection{1428,
   Author = {A. Y. Smirnov},
   Title = {Course 11 - Neutrino Mass and Mixing: Toward the Underlying Physics},
   BookTitle = {Les Houches},
   Editor = {Kazakov, Dmitri and Lavignac, Stéphane and Dalibard, Jean},
   Publisher = {Elsevier},
   Volume = {84},
   Pages = {573-653},
      Year = {2006} }

@Article{5,
   Author = {H. Tawara and T. Kato},
   Title = {Total and partial ionization cross sections of atoms and ions by electron impact},
   Journal = {Atomic Data and Nuclear Data Tables},
   Volume = {36},
   Number = {2},
   Pages = {167-353},
   Year = {1987} }

@Article{1431,
   Author = {D. R. Tilley and C. M. Cheves and J. L. Godwin and G. M. Hale and H. M. Hofmann and J. H. Kelley and C. G. Sheu and H. R. Weller},
   Title = {Energy levels of light nuclei A=5, 6, 7},
   Journal = {Nuclear Physics A},
   Volume = {708},
   Number = {1},
   Pages = {3-163},
   Year = {2002} }

@Article{1412,
   Author = {T. A. Welton},
   Title = {Some Observable Effects of the Quantum-Mechanical Fluctuations of the Electromagnetic Field},
   Journal = {Physical Review},
   Volume = {74},
   Number = {9},
   Pages = {1157-1167},
   Year = {1948} }






\end{document}